\documentclass{aa}  

\usepackage{graphicx}
\usepackage{txfonts}
\usepackage[colorlinks=true,linkcolor=blue, citecolor=blue, filecolor=blue, urlcolor=blue]{hyperref}
\bibpunct[]{(}{)}{;}{a}{}{,}

\begin{document}

   \title{Complex Organic Molecules in Protostars\\ with ALMA Spectral
Surveys (COMPASS)}

   \subtitle{VII. First interstellar detection of fully deuterated methanol}
   \titlerunning{COMPASS: First interstellar detection of fully deuterated methanol}

   \author{A. Belloche\inst{1}
        \and S. Spezzano\inst{2}
        \and A. Coutens\inst{3}
        \and M.~N. Drozdovskaya\inst{4,5}
        \and J.~K. J{\o}rgensen\inst{6}
        \and J.-E. Lee\inst{7}
        \and A.~L. Plunkett\inst{8}
        \and\\ D. Harsono\inst{9}
        \and S. Maret\inst{10}
        \and F. Cruz-S{\'a}enz de Miera\inst{3}
        \and J.-H. Jeong\inst{7}
        \and P. Nazari\inst{11}
        \and Y.-L. Yang\inst{12}
        \and H.-S. Yun\inst{7}
        \and S. Zeng\inst{12}
        }

   \institute{Max-Planck-Institut f{\"u}r Radioastronomie, 
              Auf dem H{\"u}gel 69, 53121 Bonn, Germany\\
             \email{belloche@mpifr-bonn.mpg.de}
            \and 
              Max-Planck-Institut f{\"u}r extraterrestrische Physik, 
              Giessenbachstrasse 1, 85748 Garching bei M{\"u}nchen, Germany 
            \and
              Institut de Recherche en Astrophysique et Plan\'etologie, 
              Universit\'e de Toulouse, CNRS, CNES, 9 av. du Colonel Roche, 
              31028 Toulouse Cedex 4, France
            \and
              Physikalisch-Meteorologisches Observatorium Davos und 
              Weltstrahlungszentrum (PMOD/WRC), Dorfstrasse 33, 7260 Davos 
              Dorf, Switzerland
            \and
              Department of Chemistry, Biochemistry and Pharmaceutical 
              Sciences (DCBP), Universit{\"a}t Bern, Freiestrasse 3, 
              3012 Bern, Switzerland
            \and
              Niels Bohr Institute, University of Copenhagen, Jagtvej 155A, DK 
              2200 Copenhagen N, Denmark
            \and
              Department of Physics and Astronomy, Seoul National University, 
              1 Gwanak-ro, Gwanak-gu, Seoul 08826, Republic of Korea
            \and
              National Radio Astronomy Observatory, 520 Edgemont Road, 
              Charlottesville, VA 22903, USA
            \and
              Institute of Astronomy, Department of Physics, National Tsinghua 
              University, No. 101, Section 2, Kuang-Fu Road, Hsinchu 30013, 
              Taiwan, R.O.C
            \and
              Univ. Grenoble Alpes, CNRS, IPAG, 38000 Grenoble, France
            \and
              European Southern Observatory, Karl-Schwarzschild-Strasse 2, 
              85748 Garching, Germany
            \and
              Star and Planet Formation Laboratory, RIKEN Pioneering Research 
              Institute, Wako-shi, Saitama, 351-0198, Japan
            }

   \date{Received 27 February 2026 / Accepted 26 March 2026}

  \abstract
   {There have been no robust identifications of fully deuterated methanol,
   CD$_3$OD, in the interstellar medium (ISM) to date.}
   {Our goal is continue the search for CD$_3$OD in star-forming regions to 
   determine the deuteration pattern of methanol and set constraints
   on the deuteration process.} 
   {We used a sensitive imaging spectral survey of the protostellar
   system IRAS4A in the NGC1333 molecular cloud performed as part of the large 
   program carried out with the Atacama Large Millimeter/submillimeter Array
   (ALMA) called Complex Organic Molecules in Protostars with ALMA 
   Spectral Surveys (COMPASS). The column densities of methanol and several of
   its less abundant isotopologs, including deuterated ones, were derived 
   under the assumption of local thermodynamic equilibrium (LTE).}
   {Along with the detection of several methanol isotopologs toward the
   binary component IRAS4A2, we report the first interstellar detection of 
   CD$_3$OD. The deuteration
   pattern of methanol in IRAS4A2 is similar to the pattern measured in other 
   low-mass protostars, with an increase in the apparent D/H ratio from the 
   singly to multiply deuterated isotopologs. Chemical models of prestellar 
   and protostellar cores focused on deuteration do not reproduce this pattern 
   quantitatively. This discrepancy is even more exacerbated by the detection 
   of CD$_3$OD.
   Our LTE modeling of the spectra yields low column density ratios
   of methanol to its $^{13}$C and $^{18}$O isotopologs while the $^{13}$C to 
   $^{18}$O and $^{18}$O to $^{17}$O ratios are consistent with the isotopic
   ratios expected in the local ISM. We suspect that the 
   column density of methanol is underestimated. One possible explanation is 
   that the emission is clumpy rather than monolithic.}
   {The first detection of a fully deuterated COM highlights the power of 
   deuterium fractionation as a diagnostic tool for investigating the 
   formation and inheritance of organic molecules during the process of star 
   formation.}


   \maketitle
%
%________________________________________________________________

\section{Introduction}
\label{s:introduction}
 
In the dense cores of molecular clouds, the abundance of deuterium-containing 
molecules relative to their non-deuterated forms
\citep[e.g.,][]{Ceccarelli14,Jorgensen20} can reach levels that are 
several orders of magnitude higher than the elemental abundance of deuterium 
relative to hydrogen in the Galactic disk 
\citep[$\sim$$1.6\times10^{-5}$,][]{Linsky06}. This is primarily because at the
low temperatures prevailing in these cores, the exothermicity of the reaction 
of H$_3^+$ with HD that produces H$_2$D$^+$ leads to a transfer of deuterium 
from its molecular reservoir (HD) to other molecules in the gas phase
\citep[e.g.,][]{Watson76,Herbst82}. This process is amplified in these cold 
and dense environments by the depletion of gas-phase CO onto dust grains, 
which reduces the destruction rate of H$_3^+$ and its deuterated isotopologs,
H$_2$D$^+$, D$_2$H$^+$, and D$_3^+$ 
\citep[][]{Bacmann03}. Molecules formed on the surface of dust grains also 
benefit from this enrichment in deuterium because the dissociative 
recombination of the deuterated isotopologs of H$_3^+$ liberates atomic 
deuterium that can stick to the grain surface \citep[][]{Roberts03}. 
Additionally, the ortho-to-para ratio of H$_2$ plays a role because the higher 
internal energy of ortho H$_2$ compared to para H$_2$ allows for the former to 
react with H$_2$D$^+$ even at low temperatures, implying that a high 
ortho-to-para ratio of
H$_2$ has a negative impact on deuteration \citep[][]{Flower06}. As the 
ortho-to-para ratio of H$_2$ decreases during the evolution of a molecular 
cloud \citep[][]{Flower84} and the temperature and CO abundance in dense cores 
decrease with time as well \citep[][]{Zucconi01,Bacmann02}, the efficiency of 
deuterium fractionation increases with time during the prestellar phase of a 
dense core. The deuterium fractionation level of a 
molecule\footnote{Observationally, we define the level of deuterium 
fractionation of a molecule as the column density ratio of the D-bearing 
isotopolog to the H-bearing one.} can thus reflect the time when it was 
formed. As a result, deuteration can inform us about the history of a 
star-forming region. For instance, there is evidence that deuterated water 
forms at a later stage than water during the evolution of a molecular 
cloud \citep[e.g.,][]{Taquet13,Furuya16}.

Methanol (CH$_3$OH) plays a central role in astrochemistry because it is
a key species for grain surface reactions that lead to the formation of more 
complex molecules. This species is the most abundant complex organic molecule 
(COM\footnote{Complex organic molecules are defined as carbon-bearing 
molecules that contain at least six atoms \citep[][]{Herbst09}.}) in 
star-forming regions and is often taken as a reference when deriving the 
chemical composition of such regions. Its main formation route is the 
hydrogenation of CO in the ice mantles of dust grains during the prestellar 
phase \citep[e.g.,][]{Charnley97,Watanabe02,Garrod06,JimenezSerra25}. 
Several deuterated isotopologs of methanol have been identified in the 
interstellar medium (ISM), initially with single-dish telescopes: the singly 
deuterated isotopologs CH$_3$OD and CH$_2$DOH were first detected toward the 
high-mass star forming region Orion~IRc2 \citep[][]{Mauersberger88,Jacq93},
while the multiply deuterated isotopologs CHD$_2$OH and CD$_3$OH were later 
detected in the low-mass protostellar system IRAS~16293--2422 
\citep[hereafter IRAS16293;][]{Parise02,Parise06}. Since these early 
detections, all four isotopologs have been reported in other regions or on 
smaller scales; for instance, toward both components of the binary system 
IRAS16293 \citep[][]{Jorgensen18,Manigand20,Drozdovskaya22,Ilyushin22} or 
toward the low-mass protostars HOPS 373SW and BHR71-IRS1 
\citep[][]{Lee23,Coutens26}. The high level of deuteration reported in these 
studies point to the formation of methanol during the cold prestellar phase,
which is supported by the high levels of methanol deuteration that were
measured and modeled in prestellar cores 
\citep[][]{Lin23,Spezzano26,Riedel23,Riedel25}. In addition, under the 
assumption of a purely statistical distribution of deuterium and hydrogen, 
these protostellar and prestellar studies show that the D/H ratio inferred from
each isotopolog increases with the number of deuterium atoms in the molecule 
\citep[see, e.g.,][]{Coutens26,Lin23}. 

While the latter observational result could suggest that multiply deuterated 
isotopologs formed later than singly deuterated ones during the 
prestellar phase, it might also indicate that the assumption of a purely
statistical distribution of D and H is not valid and that some
kinetic or thermodynamic properties of the chemical network play a specific
role in driving the deuteration process and setting the balance between the 
deuterated isotopologs. For instance, it was experimentally shown that 
abstraction and substitution reactions of H and D in interstellar ice 
analogs lead to deuterium enrichment of methanol, with D atoms being less 
prone to abstraction than H atoms once incorporated in the molecule 
\citep[][]{Nagaoka05}. These experiments also showed that abstraction and
substitution reactions favor the deuteration of the methyl group of methanol
over the hydroxyl group, which could explain why the CH$_2$DOH/CH$_3$OD ratio
is higher than the statistical value of 3 in low- and 
intermediate-mass protostars \citep[][]{Parise06,Ratajczak11}. However, 
CH$_2$DOH/CH$_3$OD ratios around unity were reported for high-mass star 
forming regions \citep[][]{Jacq93,Ratajczak11,Belloche25}. \citet{Faure15} 
proposed that such variations in the CH$_2$DOH/CH$_3$OD ratio could be due to 
H and D exchanges between water and methanol in the ices during the warm-up 
phase.

Measuring the abundance of all deuterated isotopologs of methanol in 
star-forming regions and comparing to predictions of chemical models could 
help us identify which reactions are key in the deuteration process. 
Unfortunately, little spectroscopic information is currently 
available for the multiply deuterated isotopologs CH$_2$DOD and CHD$_2$OD
\citep[e.g.,][]{Su91a,Su91b} and no spectroscopic predictions are 
available in public molecular databases for these two isotopologs, thereby 
preventing their identification in the ISM. However, the spectroscopy of fully 
deuterated methanol (CD$_3$OD) was investigated in the laboratory and 
spectroscopic predictions suitable for an interstellar search in the 
submillimeter (submm) wavelength range are available \citep[][]{Ilyushin23}.
Hints of the presence of this isotopolog in IRAS16293 and BHR71-IRS1 were 
reported by \citet{Ilyushin23} and \citet{Coutens26}. A robust identification 
of this isotopolog is still lacking, however. Here, we take advantage
of the large program performed with the Atacama Large Millimeter/submillimeter
Array (ALMA) called Complex Organic Molecules in Protostars with ALMA 
Spectral Surveys \citep[COMPASS;][]{Jorgensen26} to continue the search for 
CD$_3$OD. COMPASS consists in a sensitive imaging spectral survey of 11 
low-mass protostellar systems harboring hot corinos, which are hot and compact 
regions that are bright in rotational emission of COMs. In this work, we 
focus our search for CD$_3$OD on the protostellar binary system IRAS4A in 
the NGC~1333 molecular cloud at a distance of 290~pc \citep[see][ for a brief 
overview of the properties of IRAS4A]{Jorgensen26}. It is the first 
protostellar environment where fully deuterated ammonia (ND$_3$) was 
identified \citep[][]{vanderTak02}.

The article is structured as follows. We present the observatonal setup and
the method used to extract column densities from the spectra in 
Sect.~\ref{s:observations}. The results are reported in Sect.~\ref{s:results} 
and discussed in Sect.~\ref{s:discussion}. We state the conclusions of this
study in Sect.~\ref{s:conclusions}.

\section{Observations and radiative transfer modeling}
\label{s:observations}

\subsection{ALMA observations}
\label{ss:alma}

IRAS4A was observed with ALMA in the framework of the COMPASS large program
\citep[][]{Jorgensen26}\footnote{https://erda.ku.dk/vgrid/COMPASS/} 
between 2023 January and 2024 August. Details on the observational setup, 
data reduction, and continuum determination and subtraction can 
be found in \citet{Plunkett26}. The COMPASS data reduction pipeline is 
archived on Zenodo \citep[][]{Maret25}. In short, the source was observed with 
nine different setups, delivering a seamless spectral survey between 279.0 and 
311.7~GHz with a spectral resolution of 488~kHz (0.52--0.47~km~s$^{-1}$) and
a channel spacing of 244~kHz. The nine setups, which we name here S1--S9, 
have four slightly overlapping spectral windows each, labeled W25, W27, W29, 
and W31. The angular resolution (half-power beam width, HPBW) depends on the 
setup and spectral window. It varies between $0.35\arcsec\times0.20\arcsec$ 
(S6W31) and $0.62\arcsec\times0.37\arcsec$ (S1W25), with median values of 
$0.39\arcsec\times0.23\arcsec$. The rms noise level per channel ranges from 
1.6 (S4W27) to 3.3~mJy~beam$^{-1}$ (S5W25), with a median value of 
2.5~mJy~beam$^{-1}$.

\subsection{Radiative transfer modeling}
\label{ss:weeds}

We analyzed and modeled the spectra under the assumption of local thermodynamic
equilibrium \citep[LTE; see][ for a justification of this approximation for 
the COMPASS sources]{Coutens26} with Weeds \citep[][]{Maret11}, an extension 
of CLASS which is part of the GILDAS suite of 
softwares\footnote{https://www.iram.fr/IRAMFR/GILDAS}. The formalism used by
Weeds is similar to the one described by \citet{Coutens26}. Weeds computes the
radiative transfer along the line of sight for each transition of a given 
molecule, including the contribution of a continuum background, taken here to 
be the continuum level in the COMPASS continuum image in the spectral window 
of the modeled transition. Weeds accounts for the finite angular resolution of 
the observations. We modeled the emission of each considered molecule 
separately, adding 
their respective contributions together in the end. This approach is adequate 
for optically thin emission. It is also fine for optically thick emission of 
molecules that emit along distinct lines of sight within the beam. However, it 
overestimates the signal when several molecules share the same volume or are 
distributed at different distances along the same line of sight and the 
emission of at least one of them is optically thick. 

The LTE model of each molecule is defined by five free parameters: the 
emission size, $\theta_s$ (full width at half maximum, FWHM), column density, 
$N$, rotational temperature, $T$, velocity offset, $V_{\rm off}$, with respect 
to the assumed velocity of the source, and FWHM line width, $\Delta V$. For 
a given molecule, the synthetic rotational spectra of different vibrational 
states were computed independently with the same set of LTE parameters as that
used for the ground vibrational state 
\citep[see Sect.~2.2 of][ for more details on this]{Belloche25}.

\subsection{Spectroscopy}
\label{ss:spectro}

To model the spectra of methanol and its isotopologs, we used 
spectroscopic entries publicly available in the Cologne Database for Molecular 
Spectroscopy 
\citep[CDMS\footnote{https://cdms.astro.uni-koeln.de},][]{Mueller01,Mueller05}
for all but one isotopologs. 
These entries are mainly based on \citet{Xu08} for CH$_3$OH, \citet{Xu97} for
$^{13}$CH$_3$OH, \citet{Fisher07} and \citet{Sastry96} for CH$_3$$^{18}$OH,
\citet{Mueller24} for CH$_3$$^{17}$OH, \citet{Ilyushin24} for CH$_3$OD,
\citet{Coudert21} and \citet{Drozdovskaya22} for CHD$_2$OH, \citet{Ilyushin22} 
for CD$_3$OH, and \citet{Ilyushin23} for CD$_3$OD. For CH$_2$DOH, we used a 
private entry partly based on \citet{Coudert14} (L. Coudert, priv. comm.).
The partition functions of all these entries already include the torsional
contribution; thus, we did not apply any vibrational corrections to 
the column densities.

\section{Results}
\label{s:results}

\subsection{Derivation of LTE parameters}
\label{ss:param}

We analyzed the COMPASS spectrum at the peak of the channel count map 
presented in \citet{Jorgensen26} toward IRAS4A, which shows the regions rich 
in molecular line emission, excluding simple molecules. The ICRS
equatorial coordinates of this peak position are
($03^{\rm h}29^{\rm m}10\fs4329$, $31^\circ13\arcmin31\farcs97$). It 
is only 0.07$\arcsec$ away from the COMPASS peak of continuum emission of the 
binary component IRAS4A2 located at ($03^{\rm h}29^{\rm m}10\fs4375$, 
$31^\circ13\arcmin32\farcs00$). Using entries downloaded from the CDMS, 
the Lille Spectroscopic Database 
\citep[LSD\footnote{https://lsd.univ-lille.fr},][]{Motiyenko25},
and the database for molecular spectroscopy of the Jet Propulsion Laboratory  
\citep[JPL\footnote{https://spec.jpl.nasa.gov},][]{Pickett98}, as well as
several private spectroscopic entries used in \citet{Belloche25}, we obtained a
preliminary model of the molecular emission of IRAS4A2. This model currently
contains the contribution of 154 catalog entries.

Figures~B.1--B.14 show the best-fit synthetic spectra (see below for details 
on how the fits were performed) obtained for 
CH$_3$OH $\varv_{\rm t}=0, 1, 2$, $^{13}$CH$_3$OH $\varv_{\rm t}=0, 1$,
CH$_3$$^{18}$OH $\varv_{\rm t}=0, 1$,
CH$_3$$^{17}$OH $\varv_{\rm t}=0$, 
CH$_2$DOH $\varv_{\rm t}=0$,
CH$_3$OD $\varv_{\rm t}=0, 1$, 
CHD$_2$OH $\varv_{\rm t}=0$,
and CD$_3$OH $\varv_{\rm t}=0, 1$, respectively, along with the spectrum of our 
preliminary model that includes all species identified so far, overlaid on the 
observed COMPASS spectrum of IRAS4A2. The figures display all the relevant 
transitions of these
isotopologs. Transitions that were predicted to be fainter than three times the 
noise level or severely blended with emission from other (identified 
or unidentified) species are not shown. 

\begin{figure}[!ht]
\resizebox{0.99\hsize}{!}{\includegraphics[angle=0]{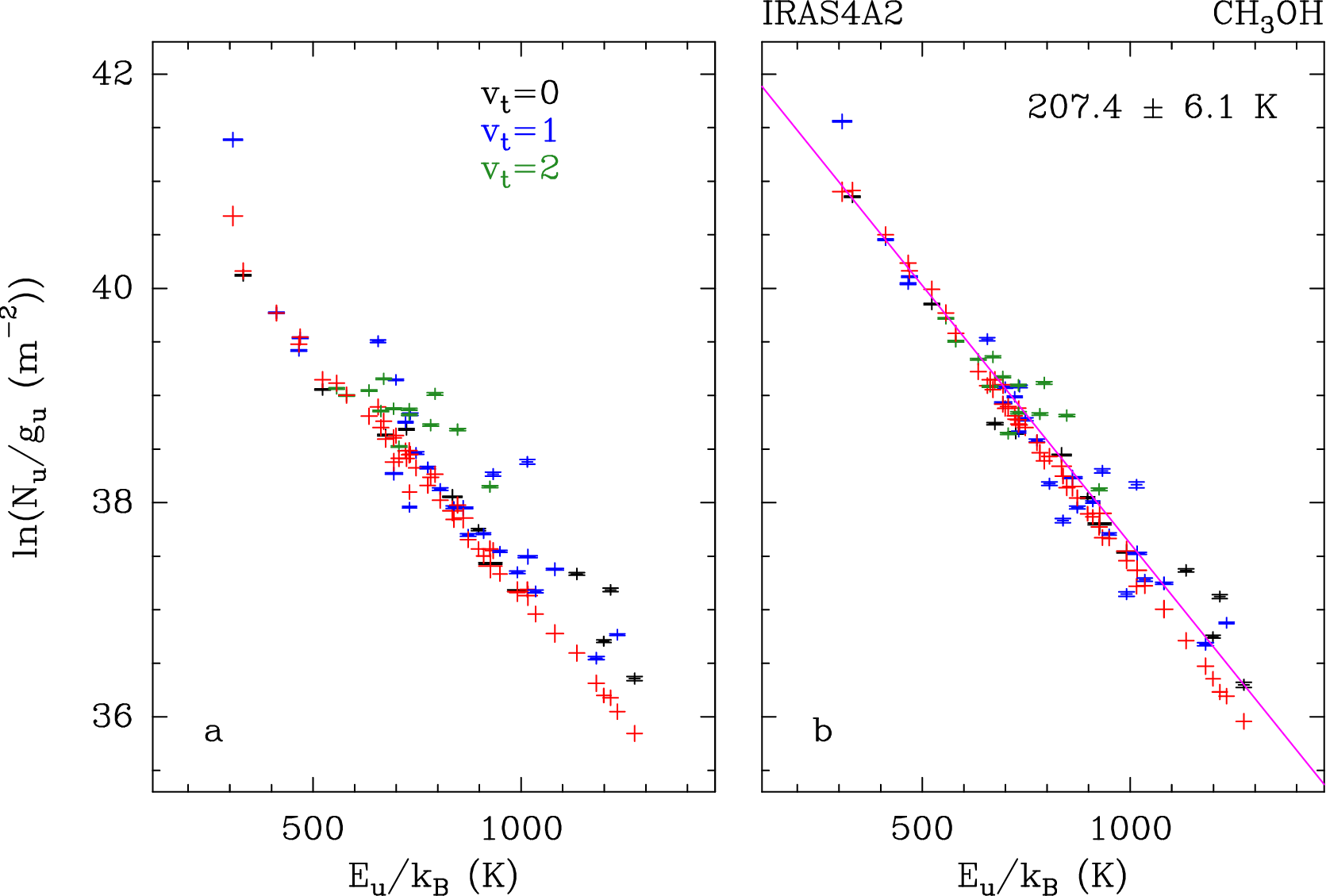}}\\[0.4ex]
\resizebox{\hsize}{!}{\includegraphics[angle=0]{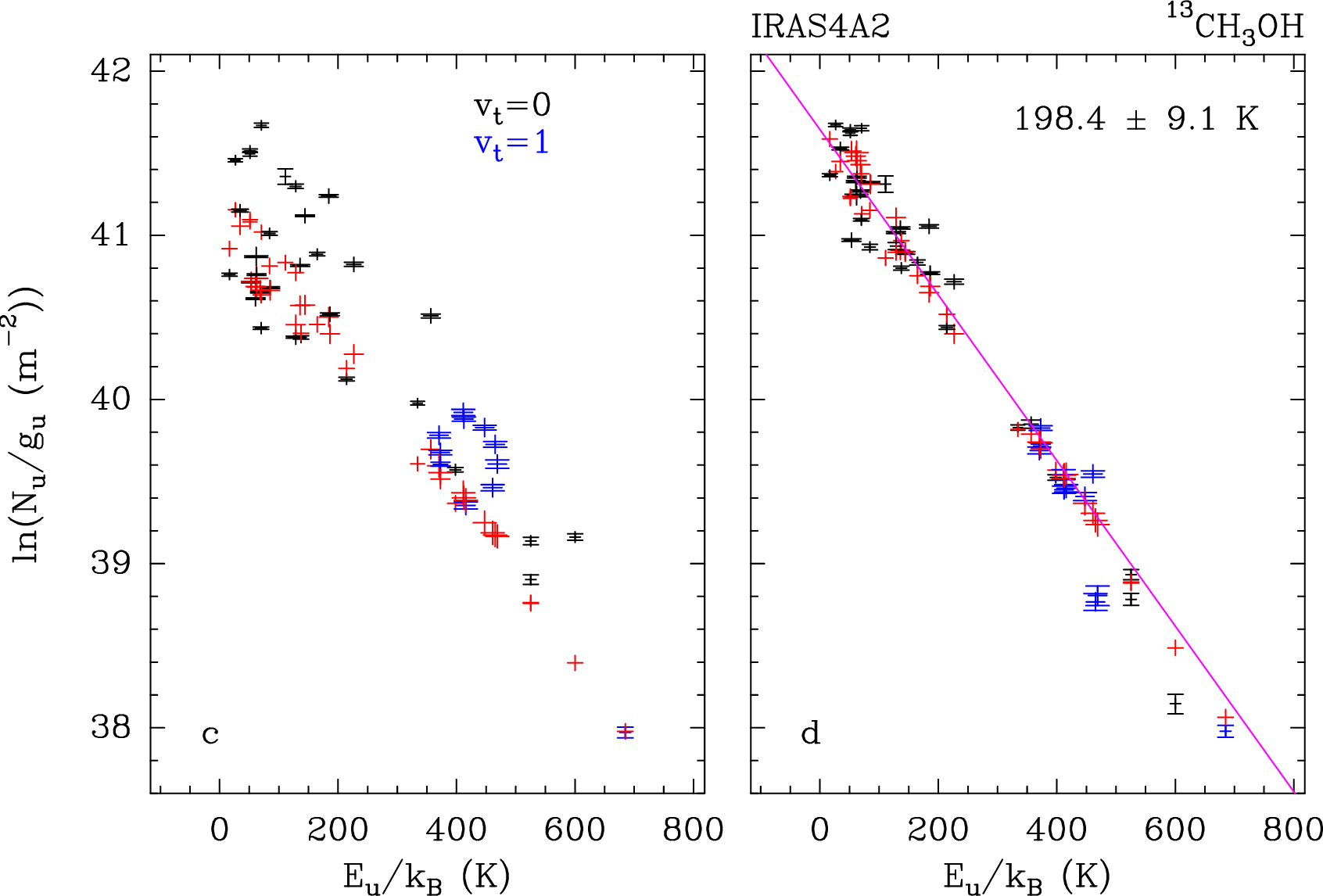}}\\[0.7ex]
\resizebox{0.99\hsize}{!}{\includegraphics[angle=0]{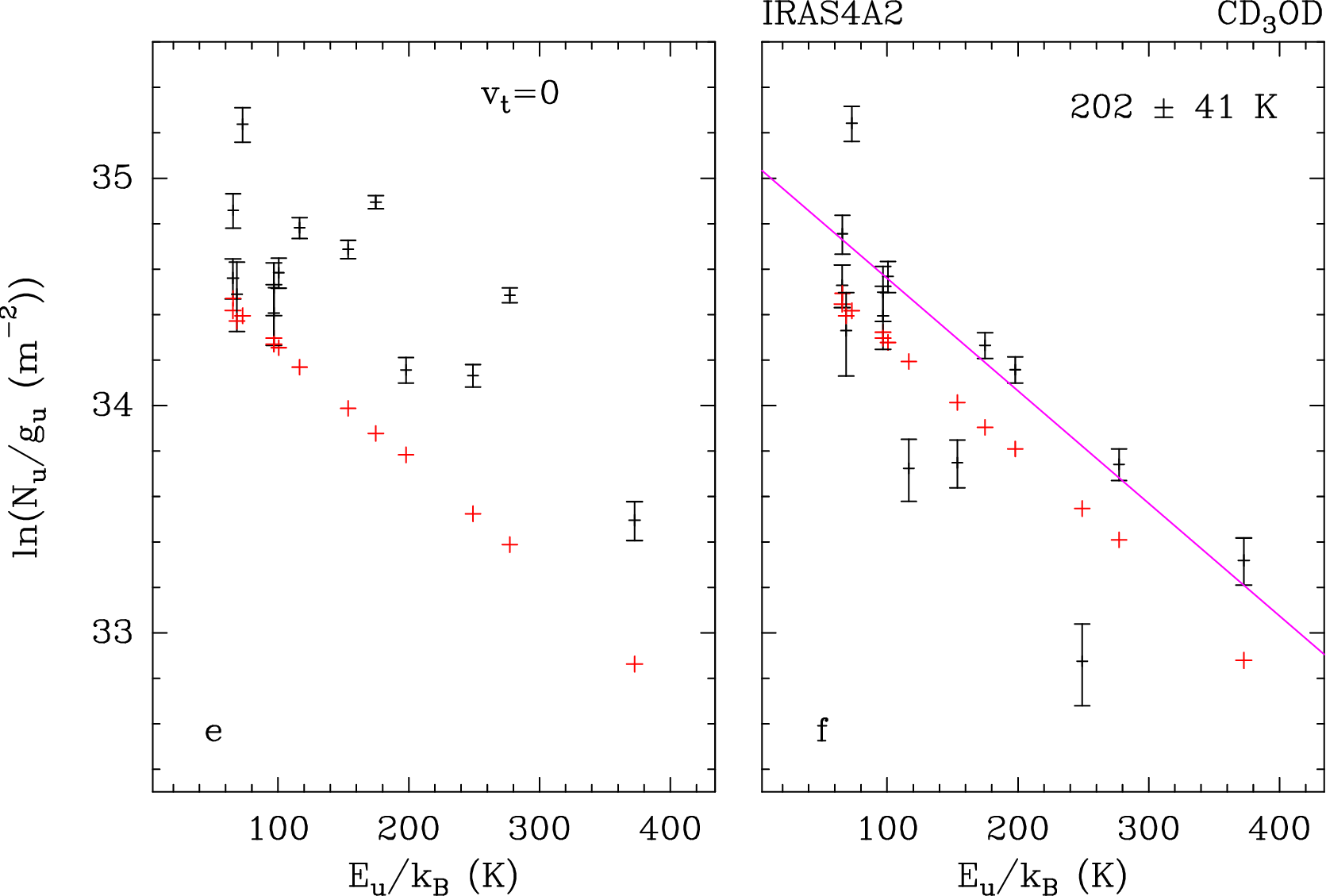}}
\caption{Population diagrams of CH$_3$OH (a and b), $^{13}$CH$_3$OH (c and d),
and CD$_3$OD (e and f) toward IRAS4A2. The observed data points are shown in 
various colors (corresponding to the torsional states) except for red as 
indicated in the upper right corner of panels a, c, and e while the synthetic 
populations of the best-fit Weeds model are shown in red. No correction is 
applied in panels a, c, and e. In panels b, d, and f, the optical depth 
correction has been applied to both the observed and synthetic populations and 
the contamination by all other species included in the full model has been 
removed from the observed data points. The purple line is a linear fit to the 
observed populations (in linear-logarithmic space) and the rotational 
temperature derived from this fit is indicated in the top right corner.}
\label{f:popdiags}
\end{figure}

\begin{figure}[!ht]
\centerline{\resizebox{1.0\hsize}{!}{\includegraphics[angle=0]{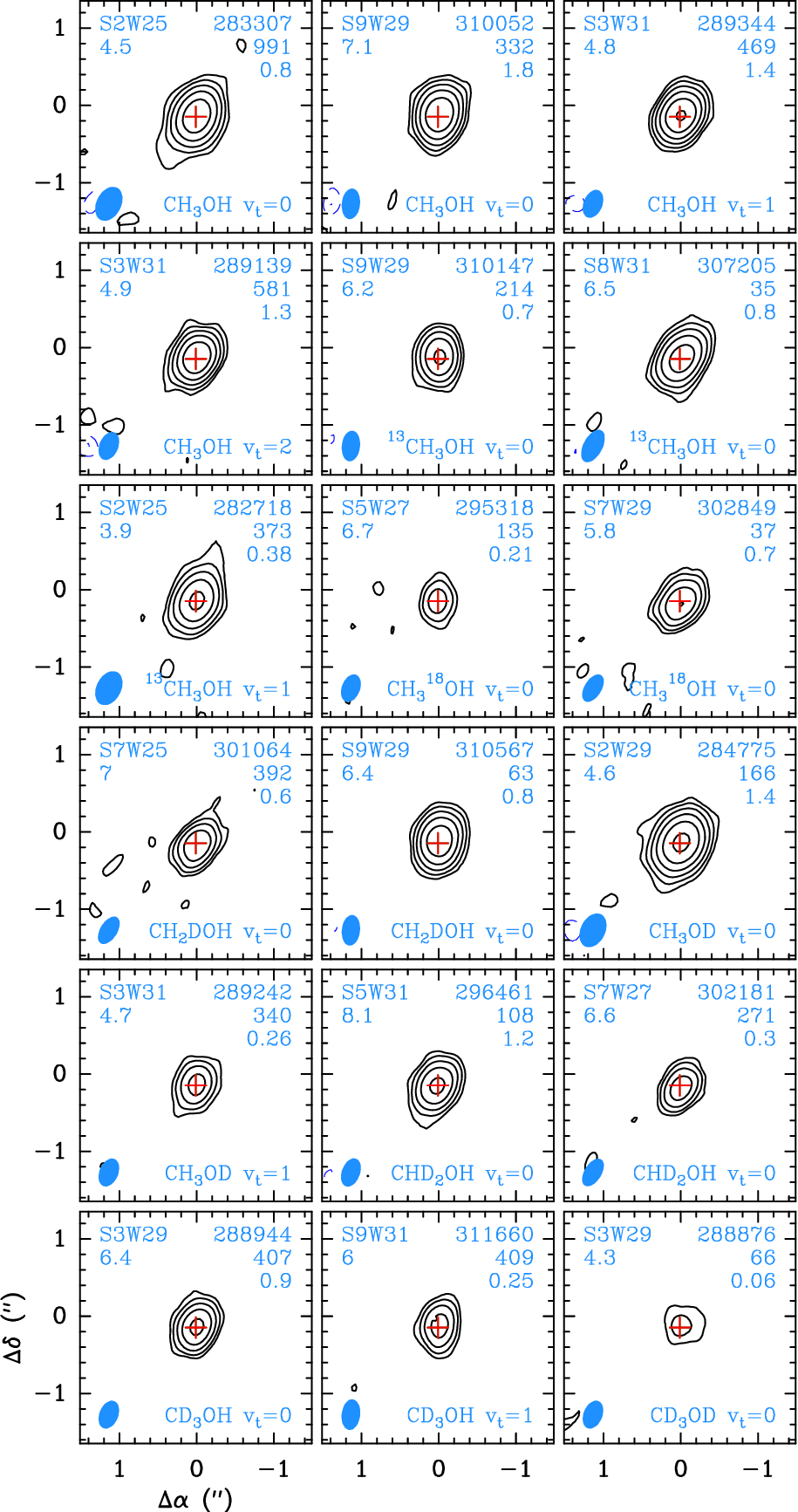}}}
\caption{Integrated intensity maps of selected transitions of methanol
isotopologs detected toward IRAS4A2. The equatorial 
offsets are defined with respect to the phase center of the observations. In 
each panel, the black contours start at 3$\sigma$ with $\sigma$ the noise 
level of the integrated intensity map. They increase by a factor of two at each 
step. Negative levels at $-3\sigma$ are represented by dashed blue contours. 
The red cross indicates the peak position of the channel count map. The setup 
(S) and its spectral window (W) are given in the top left corner of each 
panel, the noise level $\sigma$ is written below (in mJy/beam~km~s$^{-1}$), and 
the corresponding HPBW is shown as a blue ellipse in the 
bottom left corner. The frequency (in MHz), upper-level energy (in K), and 
peak opacity of the selected transition in the best-fit LTE model
are indicated from top to bottom in the top right corner. The molecule and 
vibrational state of the selected transition are indicated at the bottom.}
\label{f:maps}
\end{figure}

The best-fit models of the methanol isotopologs were obtained in several steps, 
following the same strategy as in \citet{Belloche20,Belloche25}, which we 
describe in the rest of this section. We estimated the rotational temperatures
with population diagrams using only transitions that have an opacity lower
than 2 in the final models and are not much contaminated by other species 
according to our full Weeds model. These diagrams also provide a concise 
overview of the agreement between synthetic and observed spectra for each 
methanol isotopolog. The presence of several temperature components would for
instance be easy to notice in these diagrams. The population diagrams are 
presented in Figs.~\ref{f:popdiags} and 
\ref{f:popdiag_ch3oh_18o}--\ref{f:popdiag_cd3oh}. In the right panel of each of 
these figures, we subtracted the contamination by other species on the basis 
of the full Weeds model and applied the correction for optical depth defined by 
\citet{Goldsmith99}. A weighted fit to these corrected diagrams yields a 
rotational 
temperature on the order of 200~K for all isotopologs but one, CH$_3$$^{17}$OH. 
The fitted temperatures and their statistical uncertainties are listed in 
Table~\ref{t:popfit}. As can be seen in Fig.~\ref{f:popdiag_ch3oh_17o}, the 
fit to the population diagram of CH$_3$$^{17}$OH is uncertain and appears to
be biased by two bright data points with small statistical uncertainties 
(transitions at $\sim$283.62 and 306.91~GHz, which are the uppermost data 
points in the right panel of Fig.~\ref{f:popdiag_ch3oh_17o}) that suffer from 
contamination from unidentified species in their low-frequency wing (see 
corresponding panels in Fig.~B.8). Apart from these two 
lines, the model with 200~K (red crosses in Fig.~\ref{f:popdiag_ch3oh_17o}) 
reproduces the CH$_3$$^{17}$OH emission well. Therefore, we adopted a 
temperature of 200~K to model the emission of all methanol isotopologs.

\begin{table}
 \begin{center}
 \caption{
 Rotational temperature of methanol isotopologs derived from their population diagrams toward IRAS4A2.
}
 \label{t:popfit}
 \vspace*{0.0ex}
 \begin{tabular}{lll}
 \hline\hline
 \multicolumn{1}{c}{Molecule} & \multicolumn{1}{c}{States\tablefootmark{(a)}} & \multicolumn{1}{c}{$T_{\rm fit}$\tablefootmark{(b)}} \\ 
  & & \multicolumn{1}{c}{\small (K)} \\ 
 \hline
CH$_3$OH & $\varv_{\rm t}=0$, $\varv_{\rm t}=1$, $\varv_{\rm t}=2$ & 207.4 (6.1) \\ 
\hline 
$^{13}$CH$_3$OH & $\varv_{\rm t}=0$, $\varv_{\rm t}=1$ & 198.4 (9.1) \\ 
CH$_3$$^{18}$OH & $\varv_{\rm t}=0$, $\varv_{\rm t}=1$ &   190 \hspace*{1.8ex}(42) \\ 
CH$_3$$^{17}$OH & $\varv_{\rm t}=0$ &   122 \hspace*{1.8ex}(44) \\ 
\hline 
CH$_2$DOH & $\varv_{\rm t}=0$ & 186.8 (7.0) \\ 
CH$_3$OD & $\varv_{\rm t}=0$, $\varv_{\rm t}=1$ & 197.2 (7.2) \\ 
CHD$_2$OH & $\varv_{\rm t}=0$ &   193 \hspace*{1.8ex}(13) \\ 
CD$_3$OH & $\varv_{\rm t}=0$, $\varv_{\rm t}=1$ & 197.5 (5.0) \\ 
CD$_3$OD & $\varv_{\rm t}=0$ &   202 \hspace*{1.8ex}(41) \\ 
\hline 
 \end{tabular}
 \end{center}
 \vspace*{-2.5ex}
 \tablefoot{
 \tablefoottext{a}{Torsional states that were taken into account to fit the population diagram.}
 \tablefoottext{b}{The standard deviation of the fit is given in parentheses. As explained in Sect.~3 of \citet{Belloche16} and in Sect.~4.4 of \citet{Belloche19}, this uncertainty is purely statistical and should be viewed with caution. It may underestimate the actual uncertainty.}
 }
 \end{table}

\begin{figure*}[!ht]
\centerline{\resizebox{1.0\hsize}{!}{\includegraphics[angle=0]{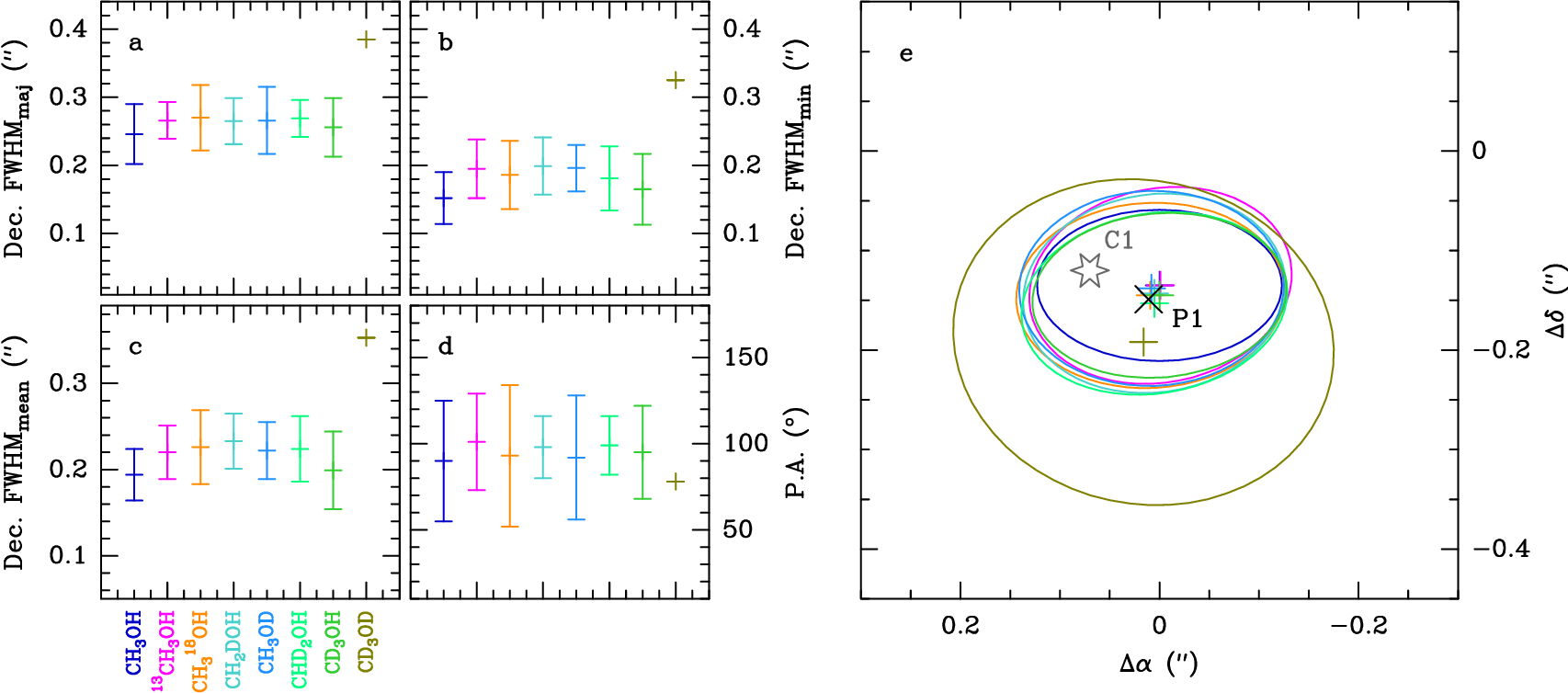}}}
\caption{Deconvolved major axis (a), minor axis (b), average size (c), and
position angle (d) of the emission of methanol and seven of its isotopologs.
For each isotopolog, the data point and its error bar represent the median and 
rms dispersion, respectively, of the measurements displayed in 
Fig.~\ref{f:sizes}. As described in Sect.~\ref{ss:param}, the deconvolved 
parameters of CD$_3$OD rely on only one faint transition and are highly 
uncertain. The deconvolved sizes are plotted as ellipses in panel e
with the same colors as in panels a--d. The plus symbols represent the median
fitted peak positions. The cross labeled P1 marks the peak position of the 
channel count map and the star symbol labeled C1 indicates the position of the 
continuum peak of IRAS4A2. The axes are labeled in equatorial offsets from the 
phase center. The deconvolved sizes are smaller than the beam (median beam 
sizes of the COMPASS data of IRAS4A: $0.39\arcsec \times 0.23\arcsec$, see
Sect.~\ref{ss:alma}).}
\label{f:sizessumm}
\end{figure*}

The emission size of each methanol isotopolog was estimated (when possible) by 
fitting two-dimensional (2D) Gaussian functions to the integrated intensity 
maps of transitions that are not contaminated by other species (according to 
our full Weeds model) and display an opacity lower than 2 in the final model 
of the isotopolog. Examples of 
such integrated intensity maps are shown in Fig.~\ref{f:maps}. The fitted 
Gaussian functions were deconvolved from the beam and the deconvolved sizes 
are displayed in Fig.~\ref{f:sizes}. A summary of all measurements is 
presented in Fig.~\ref{f:sizessumm}, where panels a--d show the median 
deconvolved sizes (major, minor, and geometrical mean) and position angle for 
each isotopolog. Panel e displays the median sizes and orientation in the 
plane of the sky. Apart from CD$_3$OD which is more uncertain due to the 
limited number of transitions and their relative weakness (only two could be 
fitted and only one properly deconvolved), the deconvolved position angles and 
sizes agree for all isotopologs, with values on the order of 90--100$^\circ$ 
east from the north and 0.20--0.22$\arcsec$, respectively. The second column of 
Fig.~\ref{f:sizes}, which plots the deconvolved sizes as a function of line 
opacity, indicates that lines with an opacity in the range of 1--2 tend to have 
slightly larger sizes than the more optically thin ones. This is the reason 
why we decided to use a slightly smaller size of 0.18$\arcsec$ to model the 
spectra with Weeds. Finally, the line width and velocity are straightforward 
to derive from optically thin lines. A FWHM of 2.0~km~s$^{-1}$ and velocity of 
6.8~km~s$^{-1}$ fit all isotopologs well. This was verified visually and is
illustrated in Figs.~B.1--B.14.

As a result of the analysis described above, the only parameter of each 
methanol isotopolog that remains free in the fitting of LTE synthetic spectra 
to the COMPASS data of IRAS4A2 is its column density, which we optimized 
visually. The best-fit column density corresponds to the maximum column 
density for which the associated synthetic spectrum does not overestimate the 
peak temperature of any transition of the modeled isotopolog in the COMPASS 
frequency range. There are exceptions to this strict criterion. For instance, 
lines with a high optical depth cannot be properly modeled with Weeds because 
potential excitation temperature gradients along the line of sight are not 
accounted for. The more optically thick a line is, the more it traces outer 
layers with lower densities and kinetic temperatures, both resulting in lower 
excitation temperatures and thus line saturation at a lower intensity level 
(self-absorption). Another exception is related to the uncertainty of the 
baseline level. A faint line may be overestimated by the model if the baseline 
level around that line has been overestimated due to spectral confusion. 
Examples of such cases can be seen for instance in the panels centered on
289670~MHz and 299510~MHz of Fig.~B.6.

\begin{table*}[!ht]
 \begin{center}
 \caption{
 Parameters of our best-fit LTE model of methanol and its isotopologs toward IRAS4A2.
}
 \label{t:coldens}
 \vspace*{-1.2ex}
 \begin{tabular}{lcrccccccc}
 \hline\hline
 \multicolumn{1}{c}{Molecule, tors. state} & \multicolumn{1}{c}{Status\tablefootmark{a}} & \multicolumn{1}{c}{$N_{\rm det}$\tablefootmark{b}} & \multicolumn{1}{c}{$N$\tablefootmark{c}} & \multicolumn{1}{c}{$\frac{N_{12}}{N}$\tablefootmark{e}} & \multicolumn{1}{c}{$\frac{N}{N_{12}}$\tablefootmark{e}} & \multicolumn{1}{c}{$\frac{68 \times N_{13}}{N}$\tablefootmark{f}} & \multicolumn{1}{c}{$\frac{N}{68 \times N_{13}}$\tablefootmark{f}} & \multicolumn{1}{c}{$\frac{560 \times N_{18}}{N}$\tablefootmark{g}} & \multicolumn{1}{c}{$\frac{N}{560 \times N_{18}}$\tablefootmark{g}} \\ 
  & & & \multicolumn{1}{c}{\small (cm$^{-2}$)} & & \multicolumn{1}{c}{\small (\%)} & & \multicolumn{1}{c}{\small (\%)} & & \multicolumn{1}{c}{\small (\%)} \\ 
 \hline
 CH$_3$OH, $\varv_{\rm t}=0$ & d & 52 &  5.0 (18)\tablefootmark{d} &       1 &  \ldots & \ldots & \ldots & \ldots & \ldots \\ 
 \hspace*{7.7ex} $\varv_{\rm t}=1$ & d & 34 &  5.0 (18)\tablefootmark{d} &       1 &  \ldots & \ldots & \ldots & \ldots & \ldots \\ 
 \hspace*{7.7ex} $\varv_{\rm t}=2$ & d & 11 &  5.0 (18)\tablefootmark{d} &       1 &  \ldots & \ldots & \ldots & \ldots & \ldots \\ 
\hline 
 $^{13}$CH$_3$OH, $\varv_{\rm t}=0$ & d & 38 &  5.5 (17) &     9.1 &  \ldots &      68 &  \ldots & \ldots & \ldots \\ 
 \hspace*{9.4ex} $\varv_{\rm t}=1$ & d & 11 &  5.5 (17) &     9.1 &  \ldots &      68 &  \ldots & \ldots & \ldots \\ 
 CH$_3$$^{18}$OH, $\varv_{\rm t}=0$ & d & 11 &  1.0 (17) &      50 &  \ldots &     370 &  \ldots &     560 &  \ldots \\ 
 \hspace*{9.4ex} $\varv_{\rm t}=1$ & t & 1 &  1.0 (17) &      50 &  \ldots &     370 &  \ldots &     560 &  \ldots \\ 
 \hspace*{9.4ex} $\varv_{\rm t}=2$ & n & 0 & $<$  2.3 (18) & $>$     2.2 &  \ldots & $>$      16 &  \ldots & $>$      24 &  \ldots \\ 
 CH$_3$$^{17}$OH, $\varv_{\rm t}=0$ & d & 4 &  2.5 (16) &     200 &  \ldots &    1500 &  \ldots &    2240 &  \ldots \\ 
\hline 
 CH$_2$DOH, $\varv_{\rm t}=0$ & d & 47 &  1.2 (18) &     4.2 &      24 &      31 &     3.2 &      47 &     2.1 \\ 
 CH$_3$OD, $\varv_{\rm t}=0$ & d & 32 &  3.0 (17) &      17 &     6.0 &     120 &    0.80 &     190 &    0.54 \\ 
 \hspace*{7.7ex} $\varv_{\rm t}=1$ & d & 14 &  3.0 (17) &      17 &     6.0 &     120 &    0.80 &     190 &    0.54 \\ 
 CHD$_2$OH, $\varv_{\rm t}=0$ & d & 76 &  1.0 (18) &     5.0 &      20 &      37 &     2.7 &      56 &     1.8 \\ 
 CD$_3$OH, $\varv_{\rm t}=0$ & d & 41 &  4.5 (17) &      11 &     9.0 &      83 &     1.2 &     120 &    0.80 \\ 
 \hspace*{7.7ex} $\varv_{\rm t}=1$ & d & 33 &  4.5 (17) &      11 &     9.0 &      83 &     1.2 &     120 &    0.80 \\ 
 CD$_3$OD, $\varv_{\rm t}=0$ & d & 10 &  2.0 (16) &     250 &    0.40 &    1870 &   0.053 &    2800 &   0.036 \\ 
 \hspace*{7.7ex} $\varv_{\rm t}=1$ & n & 0 & $<$  6.0 (16) & $>$      83 &  $<$     1.2 & $>$     620 &  $<$    0.16 & $>$     930 &  $<$    0.11 \\ 
\hline 
 \end{tabular}
 \end{center}
 \vspace*{-2.5ex}
 \tablefoot{
 All entries were modeled assuming a size (FWHM) of  0.18$\arcsec$, a temperature of  200~K, a line width (FWHM) of 2.0~km~s$^{-1}$, and a velocity of 6.8~km~s$^{-1}$.
 \tablefoottext{a}{d: detection, t: tentative detection, n: nondetection.}
 \tablefoottext{b}{Number of detected lines \citep[conservative estimate, see Sect.~3 of][ and Sect.~3.2 in this work]{Belloche16}. One line of a given species may mean a group of transitions of that species that are blended together.}
 \tablefoottext{c}{Total column density of the molecule. $x$ ($y$) means $x \times 10^y$. The partition functions of all isotopologs already include the torsional contribution, therefore no vibrational correction was applied to the column densities.}
 \tablefoottext{d}{This value likely underestimates the column density of the main isotopolog (see Sect.~\ref{ss:isoratios}).}
 \tablefoottext{e}{Column density ratio, with $N_{12}$ the column density of methanol, which is likely underestimated (see Sect.~\ref{ss:isoratios}).}
 \tablefoottext{f}{Column density ratio, with $N_{13}$ the column density of $^{13}$CH$_3$OH and 68 the assumed $\frac{^{12}{\rm C}}{^{13}{\rm C}}$ isotopic ratio.}
 \tablefoottext{g}{Column density ratio, with $N_{18}$ the column density of CH$_3$$^{18}$OH and 560 the assumed $\frac{^{16}{\rm O}}{^{18}{\rm O}}$ isotopic ratio.}
 }
 \end{table*}

\subsection{Column densities of methanol isotopologs and detection of CD$_3$OD}
\label{ss:coldens}

\begin{figure*}[!ht]
\centerline{\resizebox{0.92\hsize}{!}{\includegraphics[angle=0]{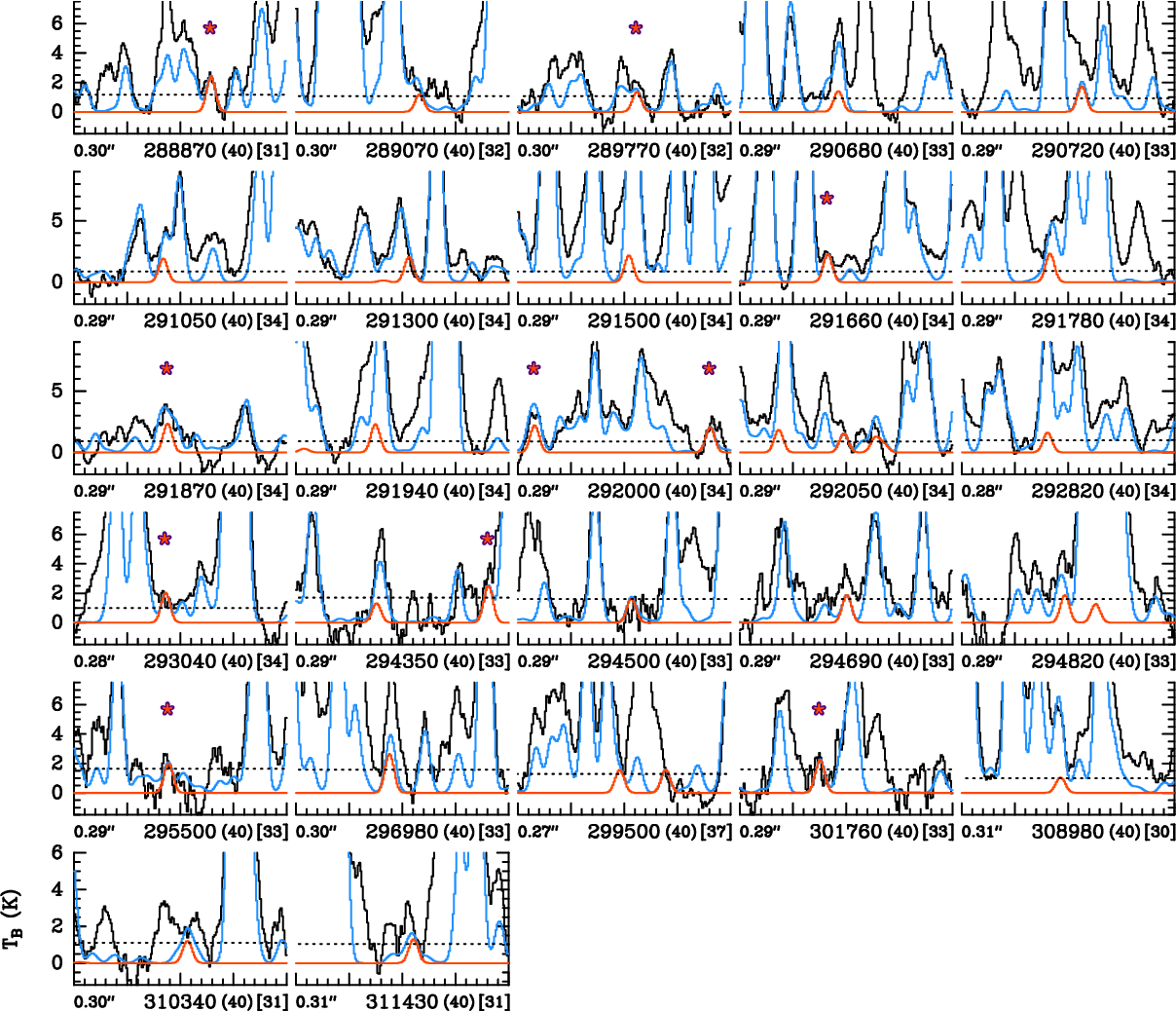}}}
\caption{Transitions of CD$_3$OD $\varv_{\rm t}=0$ covered by the COMPASS survey
toward IRAS4A2 with ALMA. The best-fit LTE synthetic spectrum of 
CD$_3$OD $\varv_{\rm t}=0$ is shown in red and overlaid on the observed 
spectrum displayed in black. The blue synthetic spectrum contains the 
contribution of all molecules included in our model of IRAS4A2 so far, 
including the molecule shown in red. The ten spectral lines of CD$_3$OD that 
have a peak signal-to-noise (S/N) ratio higher than 3 and are not 
significantly contaminated 
by emission from other (identified or unidentified) molecules, thereby firmly 
contributing to the identification of CD$_3$OD, are marked with a red star. 
The central frequency is indicated in MHz below each panel as well as the 
HPBW (geometrical mean) on the left, the width of each panel in MHz in 
parentheses, and the continuum level in K of the baseline-subtracted spectra 
in brackets. The y-axis is labeled in brightness temperature units (K). The 
dotted line indicates the 3$\sigma$ noise level.}
\label{f:spec_cd3od_ve0}
\end{figure*}

The best-fit spectra resulting from the procedure described in 
Sect.~\ref{ss:param} are overlaid on IRAS4A2's spectra in 
Fig.~\ref{f:spec_cd3od_ve0} and B.1--B.14,
and the best-fit parameters are listed in Table~\ref{t:coldens}. The number of 
lines that we consider as clearly detected (i.e., lines that have a peak 
temperature higher than $3\sigma$, coincide in frequency with a clear peak in
the observed spectrum, and are not severely blended with emission from other 
molecules) is also provided for each isotopolog. The third criterion means in
practice that we counted lines that have a modeled peak temperature 
representing at least two thirds of the detected signal, the remaining third 
being either filled by the emission of a contaminating molecule that is 
included in our preliminary full model, or still unidentified. 

All isotopologs of methanol that have spectroscopic information available in 
public databases are clearly detected toward IRAS4A2 with COMPASS: CH$_3$OH, 
$^{13}$CH$_3$OH, CH$_3$$^{18}$OH, CH$_3$$^{17}$OH, CH$_2$DOH, CH$_3$OD, 
CHD$_2$OH, CD$_3$OH, and CD$_3$OD. In particular, we report the first robust 
interstellar detection of fully deuterated methanol, CD$_3$OD, with, following 
the criteria listed above, ten lines clearly detected (see lines marked with 
red stars in Fig.~\ref{f:spec_cd3od_ve0} and listed in 
Table~\ref{t:cd3od_det_int}). Two of these ten lines are partially contaminated
by CH$_3$OCHO $\varv_{\rm t}=1$ and CH$_3$OCH$_3$ $\varv_{11}=1$ for the line at 
291867~MHz and by g-(CH$_2$OH)$_2$ $\varv=0$ and CHD$_2$OH $\varv=0$ for the 
line at 291983~MHz. 
Our preliminary LTE models of the other sources indicate that CD$_3$OD is not
detected in the rest of the COMPASS sample.

\subsection{Isotopic ratios and column density of methanol}
\label{ss:isoratios}

\begin{table*}[!t]
 \begin{center}
 \caption{
 Deuteration level of methanol toward IRAS4A2.
}
 \label{t:deuteration}
 \vspace*{-1.2ex}
 \begin{tabular}{lcccc}
 \hline\hline
 \multicolumn{1}{c}{Molecules} & \multicolumn{2}{c}{$R$ (\%)\tablefootmark{a}} & \multicolumn{2}{c}{D/H (\%)\tablefootmark{b}} \\ 
   & \multicolumn{1}{c}{$^{13}$C\tablefootmark{c}} & \multicolumn{1}{c}{$^{18}$O\tablefootmark{d}} & \multicolumn{1}{c}{$^{13}$C\tablefootmark{c}} & \multicolumn{1}{c}{$^{18}$O\tablefootmark{d}}  \\ 
 \hline
  & \multicolumn{4}{c}{Deuteration of hydroxyl group} \\ 
 \hline
 CH$_3$OD/CH$_3$OH &   0.80 $\pm$    0.20 &    0.54 $\pm$    0.21 &    0.80 $\pm$    0.20 &    0.54 $\pm$    0.21 \\ 
 CD$_3$OD/CD$_3$OH &\multicolumn{2}{c}{    4.4 $\pm$     0.8} & \multicolumn{2}{c}{    4.4 $\pm$     0.8} \\ 
 \hline
  & \multicolumn{4}{c}{Deuteration of methyl group} \\ 
 \hline
 CH$_2$DOH/CH$_3$OH &    3.2 $\pm$     0.8 &     2.1 $\pm$     0.9 &     1.1 $\pm$     0.3 &    0.71 $\pm$    0.29 \\ 
 CHD$_2$OH/CH$_3$OH &    2.7 $\pm$     0.7 &     1.8 $\pm$     0.7 &     9.4 $\pm$     1.2 &     7.7 $\pm$     1.5 \\ 
 CD$_3$OH/CH$_3$OH &    1.2 $\pm$     0.3 &    0.80 $\pm$    0.32 &      23 $\pm$       2 &      20 $\pm$       3 \\ 
 CD$_3$OD/CH$_3$OD &\multicolumn{2}{c}{    6.7 $\pm$     1.2} & \multicolumn{2}{c}{     41 $\pm$       2} \\ 
 CHD$_2$OH/CH$_2$DOH &\multicolumn{2}{c}{     83 $\pm$      13} & \multicolumn{2}{c}{     83 $\pm$      13} \\ 
 CD$_3$OH/CHD$_2$OH &\multicolumn{2}{c}{     45 $\pm$       7} & \multicolumn{2}{c}{    135 $\pm$      20} \\ 
 \hline
  & \multicolumn{4}{c}{Deuteration of both groups} \\ 
 \hline
 CD$_3$OD/CH$_3$OH &  0.053 $\pm$   0.013 &   0.036 $\pm$   0.014 &    15.2 $\pm$     1.0 &    13.7 $\pm$     1.4 \\ 
 \hline
 \end{tabular}
 \end{center}
 \vspace*{-2.5ex}
 \tablefoot{
 \tablefoottext{a}{Column density ratio.}
 \tablefoottext{b}{D/H ratio assuming a purely statistical distribution.}
 \tablefoottext{c}{Ratio computed using 68$\times$$N$($^{13}$CH$_3$OH) as a proxy for $N$(CH$_3$OH).}
 \tablefoottext{d}{Ratio computed using 560$\times$$N$(CH$_3$$^{18}$OH) as a proxy for $N$(CH$_3$OH).}
 }
 \end{table*}

Column density ratios are provided in Table~\ref{t:coldens}. The ratio of 
methanol to its $^{13}$C  isotopolog is 9 (with an uncertainty of $\sim$15\%, 
see Sect.~\ref{ss:uncertainties}), which is surprisingly low compared 
to the $^{12}$C$/^{13}$C isotopic ratio of $\sim$68 expected at the 
Galactocentric distance of the Sun \citep[with an uncertainty of about 20\%, 
see][]{Langer93,Wilson99,Boogert00,Milam05,Jacob20,Yan23,Luo24,Brunken24}. 
Likewise, the ratio of 
methanol to its $^{18}$O isotopolog is 50 (with an uncertainty of $\sim$15\%), 
a very low value compared to the $^{16}$O$/^{18}$O isotopic ratio of $\sim$560 
expected in the local ISM \citep[with an uncertainty of about 40\%, 
see][]{Wilson99,Smith09,Persson18,Agundez19}. However, we get a column density 
ratio $^{13}$CH$_3$OH/CH$_3$$^{18}$OH of 5.5 (with an uncertainty of 
$\sim$15\%), which is only slightly smaller than the ratio of $\sim$8 expected 
in the local interstellar medium \citep[][]{Gardner81,Wilson99}. For 
CH$_3$$^{18}$OH/CH$_3$$^{17}$OH, we obtain a ratio of 4.0 (with an uncertainty 
of $\sim$15\%), which is consistent with the $^{18}$O/$^{17}$O isotopic ratio
measured in nearby molecular clouds \citep[4.11 $\pm$ 0.14 in Ophiuchus and 
4.0 $\pm 0.5$ in Taurus,][]{Wouterloot05,Ladd04} and with the ratio of 
$\sim$3.8 expected at the Galactocentric distance of the Sun from the Galactic 
gradient derived by \citet[][]{Zhang20}.

As we will discuss in Sect.~\ref{ss:disc_isoratios}, we suspect that the
column density of the main isotopolog is underestimated, by maybe up to a 
factor of 7--11 if we take the $^{13}$C or $^{18}$O isotopolog as reference. 
Therefore, in the rest of the article, we use the $^{13}$CH$_3$OH and 
CH$_3$$^{18}$OH isotopologs as proxies for the column density of the main 
isotopolog, multiplying their column densities by the $^{12}$C/$^{13}$C and 
$^{16}$O/$^{18}$O isotopic ratios expected in the local ISM, $\sim$68 and 
$\sim$560, respectively. The last four columns of Table~\ref{t:coldens} thus 
provide column density ratios that use the $^{13}$C and $^{18}$O isotopologs
as proxies for the main isotopolog of methanol.

\subsection{Uncertainties on column densities and ratios}
\label{ss:uncertainties}

As explained in \citet{Coutens26}, once four of the five LTE modeling 
parameters are fixed (emission size, rotational temperature, velocity, line 
width), the rms uncertainty (1$\sigma$) on the column densities is typically 
smaller than 10\%. It depends on the number of detected transitions and on 
the number of optically thin ones (the integrated intensity of which is 
proportional to the column density). Except for CD$_3$OD and CH$_3$$^{17}$OH, 
all isotopologs identified toward IRAS4A2 have more than a dozen transitions 
detected, many of them with high S/N ratios, implying a formal 
column density uncertainty below the 10\% level. For CD$_3$OD, about half of 
the ten detected lines have peak S/N ratios of $\sim$6; therefore, a
conservative estimate of the uncertainty on its column density is $\leq$15\%. 
On top of these 
formal uncertainties, the calibration uncertainty was estimated to be on the 
order of 10\% on the basis of the continuum flux density variations seen in the
COMPASS data sets \citep[][]{Plunkett26}. This directly translates into a
similar uncertainty on the column density of a molecule, provided that a 
sufficient number of lines used to derive its column density are optically 
thin.

However, if we account for the uncertainties on all LTE modeling parameters, 
then the uncertainties on the column densities increase. \citet{Coutens26} 
showed that an uncertainty of 15\% on the temperature, which is a conservative 
estimate for IRAS4A2 (see the dispersion of the temperatures listed in 
Table~\ref{t:popfit}), translates into column density uncertainties of 
$\sim$10--20\%. Figures~\ref{f:sizes} and \ref{f:sizessumm} suggest an 
uncertainty on the emission size of $\sim$15\%, implying column density 
uncertainties smaller than 30\% because the column density derived from 
optically thin lines is inversely proportional to the beam filling factor which
scales with the size squared when the size is much smaller than the beam. Here,
the size is a sizeable fraction of the beam. The dependence of the 
uncertainty on the size is thus shallower. Finally, we estimate an uncertainty 
of $\sim$10\% on the line width, which translates into a similar uncertainty on 
the column density for optically thin lines.

Accounting for all these uncertainties quadratically yields a column density
uncertainty of $\lesssim$40\%. However, in this work, we are mostly interested 
in column density ratios between methanol isotopologs. Apart from the 
statistical uncertainty coming from fitting the column density, the other 
uncertainties affect all isotopologs in a correlated way provided that the 
emission size and temperature are the same for all isotopologs, which we 
assume. These uncertainties thus do not contribute to the uncertainty on the
ratios. Therefore, with an uncertainty of $\sim$10\% on each column density
($\sim$15\% for CD$_3$OD), each column density ratio would have an uncertainty 
of $\sim$15\% ($\sim$18\% when CD$_3$OD is involved). However, 
the main source of uncertainty for the ratios that involve the column 
densities of the $^{13}$C or $^{18}$O isotopologs as proxies for the main one 
is the uncertainty on the assumed isotopic ratio ($\sim$20\% for 
$^{12}$C/$^{13}$C and $\sim$40\% for $^{16}$O/$^{18}$O, see 
Sect.~\ref{ss:isoratios}). In conclusion, the uncertainty on a column density 
ratio is $\sim$25\% when it involves the $^{13}$C isotopolog, $\sim$40\% when 
it involves the $^{18}$O isotopolog, $\sim$18\% for CD$_3$OD relative to 
another deuterated isotopolog, and $\sim$15\% in the other cases.

\subsection{Deuteration}
\label{ss:deuteration}

We report in Table~\ref{t:deuteration} column density ratios involving various 
pairs of deuterated isotopologs of methanol and methanol itself. We also 
provide the D/H ratios derived from these ratios, after accounting for the 
statistics of the D and H atoms 
\citep[see][]{Parise04,Faure15,Manigand19, Coutens26}. 
Table~\ref{t:deuteration} 
displays both the results computed with the column density of $^{13}$CH$_3$OH
 multiplied by 68 as a proxy for the column density of methanol and those 
obtained using the column density of CH$_3$$^{18}$OH multiplied by 560. 

\begin{figure*}[!t]
\sidecaption
\includegraphics[angle=0,width=0.67\hsize]{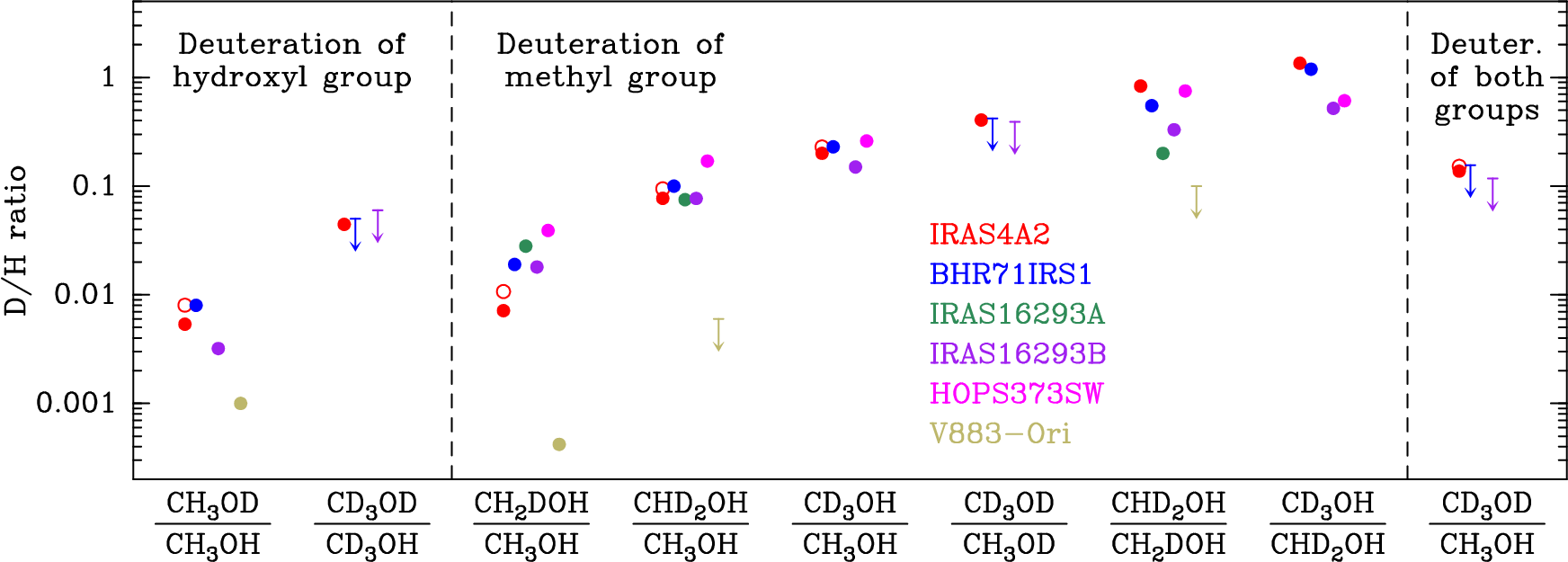}
\caption{Statistically corrected D/H ratios measured toward IRAS4A2 and other 
low-mass protostars. For IRAS4A2, the ratios that involve CH$_3$OH were 
computed using the column density of CH$_3$$^{18}$OH multiplied by the assumed 
$^{16}$O/$^{18}$O isotopic ratio (filled circle) or the column density of 
$^{13}$CH$_3$OH multiplied by the assumed $^{12}$C/$^{13}$C isotopic ratio 
(unfilled circle). The uncertainties are at most 25\% and 40\% in the former 
and latter cases, and at most 15--18\% for ratios not involving CH$_3$OH (see 
Sect.~\ref{ss:uncertainties} and Table~\ref{t:deuteration}). See 
\citet{Coutens26} for the assumptions made for the other sources.}
\label{f:deuteration}
\end{figure*}

\section{Discussion}
\label{s:discussion}

\subsection{Isotopic ratios}
\label{ss:disc_isoratios}

In our population diagram analysis of the spectra of methanol and its 
isotopologs toward IRAS4A2, we only used transitions with an opacity lower 
than 2 according to our final best-fit LTE model. All population diagrams are
consistent with a rotational temperature of $\sim$200~K and, in particular, no
deviation from a straight line is visible in the population diagram of the
main isotopolog (Fig.~\ref{f:popdiags}a), suggesting no apparent optical depth 
issue. Despite these precautions, our analysis yields a 
CH$_3$OH/$^{13}$CH$_3$OH column density ratio of $\sim$9, which is much smaller
than the typical $^{12}$C/$^{13}$C isotopic ratio in the local ISM ($\sim$68).
Similarly low values ranging from 4 to 30 have also been derived for CH$_3$OH 
and CH$_3$CN by \citet{Busch25} from NOEMA observations of low-mass protostars 
in the Solar neighborhood, including sources in the same cloud as IRAS4A2.
\citet{Busch25} interpreted the CH$_3$OH/$^{13}$CH$_3$OH column density ratios
as tracing the $^{12}$C/$^{13}$C ratio, implying that methanol is significantly
enriched in $^{13}$C. However, an argument against this interpretation of our
results is that, as mentioned in Sect.~\ref{ss:isoratios}, both the 
$^{13}$CH$_3$OH/CH$_3$$^{18}$OH and CH$_3$$^{18}$OH/CH$_3$$^{17}$OH ratios 
derived for IRAS4A2 are similar to the values expected in the local ISM
(5.5 versus $\sim$8 for the former and 4.0 versus $\sim$4 for the latter).
If methanol were enriched in $^{13}$C in IRAS4A2 (by a factor of $>$7) then it 
would have to be enriched in $^{18}$O and $^{17}$O in similar proportions, 
which sounds unlikely from a chemical point of view. It seems
more likely that our analysis of the methanol emission somehow underestimates 
the true column density of methanol. 

One could wonder whether or not our assumption of a monolithic emission over a 
size of 0.18$\arcsec$ is valid. Figure~\ref{f:sizes} shows that the 
S/N ratios of the numerous lines of methanol and its isotopologs 
detected toward IRAS4A2 are sufficiently high for a proper deconvolution of 
the emission from the beam, otherwise the scatter of the deconvolved sizes 
would be much larger. However, a clumpy methanol emission extending over
0.18$\arcsec$ would have a similar deconvolved size. Such a clumpy structure 
would mean a smaller effective beam filling factor for the radiative transfer 
along the line of sight, implying higher line opacities. The population 
diagrams shown in Figs.~\ref{f:popdiags} and 
\ref{f:popdiag_ch3oh_18o}--\ref{f:popdiag_cd3oh} would then be contaminated by 
transitions that are in fact more optically thick than what our best-fit 
models currently suggest, making it harder to derive the true column density 
of the main isotopolog for instance. The six lines with the lowest opacities 
in the population diagram of methanol shown in Fig.~\ref{f:popdiags}b have 
values in the range 0.06--0.3 in our current best-fit model. The clumpiness 
would have to be such that it increases the opacities by a factor higher than 
30 to make all the methanol transitions shown in 
Fig.~\ref{f:popdiags}b significantly optically thick ($\tau > 2$). 
Observations at higher angular resolution would be necessary to investigate the
structure of the methanol emission on scales smaller than 0.18$\arcsec$.

\subsection{Comparison to other sources}
\label{ss:comp_sources}

The deuteration levels reported in Table~\ref{t:deuteration} are shown in 
Fig.~\ref{f:deuteration} and compared to 
results reported in the literature for other low-mass protostars as compiled 
by \citet{Coutens26}. They were taken from \citet{Coutens26} for BHR71IRS1 
(continuum peak), \citet{Jorgensen18}, \citet{Manigand20}, 
\citet{Drozdovskaya22}, and \citet{Ilyushin22,Ilyushin23,Ilyushin24} for 
IRAS16293A and IRAS16293B, \citet{Lee23} for HOPS373SW, and \citet{Zeng25} and 
\citet{Jeong25} for V883-Ori. Figure~\ref{f:deuteration} reveals that the 
deuteration pattern of methanol in IRAS4A2 is almost identical to that in 
BHR71IRS1, and that it correlates well (within a factor of 3) with the D/H 
ratios of the other protostars as well, except for the eruptive object 
V883-Ori, which shows much lower ratios.

If the deuteration of methanol in each source followed a purely statistical 
distribution drawn from a fixed D/H atomic ratio, then the D/H ratios 
displayed in Fig.~\ref{f:deuteration} would lie on an horizontal line, the 
height of which could vary from source to source. This is clearly not the case 
for the sample of sources shown in Fig.~\ref{f:deuteration} (excluding 
V883-Ori), with D/H ratios ranging from $\sim$0.01 for the singly deuterated 
isotopologs (CH$_2$DOH and CH$_3$OD relative to CH$_3$OH) to values close to 
or higher than 0.1 for the multiply deuterated isotopologs (CHD$_2$OH, 
CD$_3$OH, and CD$_3$OD relative to CH$_3$OH, and CD$_3$OD relative to 
CH$_3$OD). We conclude from this that the deuteration of methanol up to its
fully deuterated form in low-mass protostars does not follow a purely 
statistical distribution drawn from a fixed D/H atomic ratio.

The D/H ratio derived from CD$_3$OD/CH$_3$OD in IRAS4A2 is similar to the
upper limits reported for BHR71IRS1 and IRAS16293B by \citet{Coutens26} and
\citet{Ilyushin23}, respectively. The data analyzed by \citet{Coutens26} and 
\citet{Ilyushin23} were not sufficient to support any claim of a robust 
interstellar detection of CD$_3$OD, but both studies reported hints of the 
presence of CD$_3$OD. In light of the clear detection in IRAS4A2, the fact 
that the methanol deuteration pattern in IRAS4A2 is nearly identical to the 
pattern 
seen in BHR71IRS1 for the less deuterated isotopologs of methanol suggests that
the upper limit reported by \citet{Coutens26} in BHR71IRS1 could be close to
the actual column density of CD$_3$OD in that source. Apart from CH$_2$DOH, the
D/H ratios of all deuterated isotopologs of methanol are lower in IRAS16293B 
than in IRAS4A2 by a factor of 1.5--3. If this trends continues for CD$_3$OD,
then its actual column density in IRAS16923B could be a factor of a few lower
than its upper limit.

\subsection{Comparison to models}
\label{ss:comp_models}

The deuteration levels of methanol reported in Table~\ref{t:deuteration} for 
IRAS4A2 correspond to the gas-phase values in the hot corino, that is the 
region where the ices have thermally sublimated from the grain surfaces. In 
this section, we compare these observational results to the predictions of 
various published models that were designed for protostellar envelopes or 
prestellar cores. The comparison to prestellar core models is motivated by the 
fact that the deuteration pattern measured in the gas phase of a hot corino 
may reflect the deuteration pattern that was achieved in the ices at the end 
of the prestellar phase before the protostar triggers thermal desorption.

\begin{figure}[!t]
\centerline{\resizebox{\hsize}{!}{\includegraphics[angle=0]{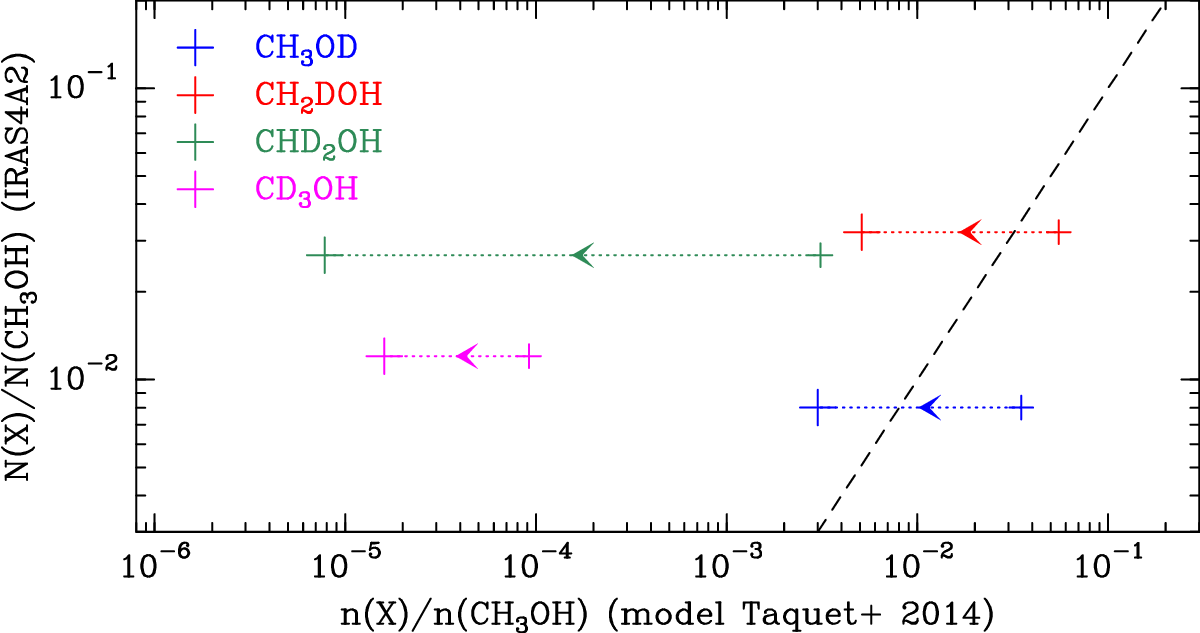}}}
\caption{Column density ratios of methanol deuterated isotopologs relative to
methanol measured toward IRAS4A2 in the gas phase (using $^{13}$CH$_3$OH as a 
proxy for methanol) versus predictions of the
model of \citet{Taquet14} at a scale of 50~au. The small and big symbols 
represent the model results at the beginning and the end of the Class~0 phase, 
respectively, and the arrows indicate the time evolution. The dashed line 
represents the one-to-one relation.}
\label{f:modelTCS14}
\end{figure}

\citet{Taquet14} presented the results of a gas-grain astrochemical model that 
follows the evolution of a collapsing dense core from the prestellar to the
protostellar phase. The strength of the model is the multilayer approach that
was adopted to describe the evolution of the ice mantles of the grains. The 
model accounts for a decrease of the ortho-to-para ratio of H$_2$ with time,
which leads to an increase of the D/H ratio of atoms that stick to the grains. 
The model also includes abstraction and substitution reactions of H and D in 
the methanol network, on the basis of measurements performed on interstellar 
ice analogs \citep[][]{Nagaoka05,Watanabe05,Nagaoka07,Hidaka09}.
Figure~\ref{f:modelTCS14} compares the column density ratios that we derived 
for the deuterated isotopologs of methanol in IRAS4A2 to the gas-phase 
abundance ratios predicted by the model of \citet{Taquet14} at a scale of 
50~au at the beginning and at the end of the Class 0 phase. The model produces 
ratios consistent with the observed ones for CH$_2$DOH and CH$_3$OD, but it 
underestimates the ratios of the multiply deuterated isotopologs CHD$_2$OH and 
CD$_3$OH by one to two orders of magnitude. While results for CD$_3$OD were 
not reported, it is likely that the model would underestimate its ratio as 
well.

\begin{figure}[!t]
\centerline{\resizebox{\hsize}{!}{\includegraphics[angle=0]{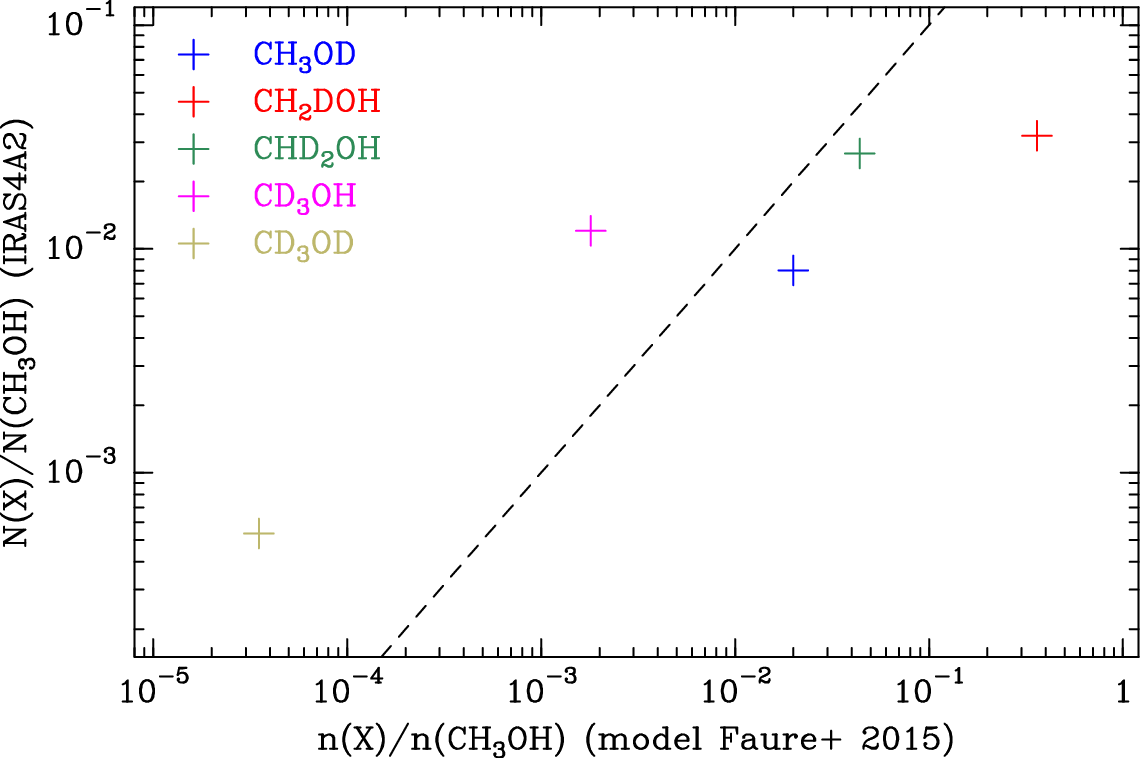}}}
\caption{Column density ratios of methanol deuterated isotopologs relative to
methanol measured toward IRAS4A2 (using $^{13}$CH$_3$OH as a 
proxy for methanol) versus predictions of the IRAS16293 best-fit 
model of \citet{Faure15}. The dashed line represents the one-to-one relation.}
\label{f:modelF15}
\end{figure}

\citet{Faure15} investigated the impact of thermally activated hydrogen 
isotope exchanges between water and methanol in interstellar ices during the 
warm-up phase that is triggered by the formation of a protostar. These 
exchanges occur only between the hydroxyl group of methanol and water 
\citep[][]{Souda03,Ratajczak09}. The model of \citet{Faure15} uses a 
simplified grain surface chemistry network. It assumes as initial conditions, 
supposed to be set during the prestellar phase, a statistical distribution of 
the deuterated isotopologs of methanol. The authors assumed a D/H ratio of 
12\% that corresponds to the CH$_2$DOH/CH$_3$OH column density ratio of 37\% 
reported by \citet{Parise06} for IRAS16293 at single-dish scales\footnote{As 
noted by \citet{Belloche16}, the column densities reported by \citet{Parise06} 
for CH$_2$DOH were overestimated by a factor of $\sim$2 because of issues with 
the spectroscopic predictions used at that time.}. The HDO/H$_2$O abundance 
ratio in the ice mantles was a free parameter of the model. 
Figure~\ref{f:modelF15} compares the results that we obtained for methanol in
IRAS4A2 to the predictions of the model of \citet{Faure15} that fitted best 
the column density ratios reported by \citet{Parise06} for IRAS16293. This 
best-fit model has an initial D/H ratio of water in the ice of 2\%. The model 
predictions are close to the one-to-one relation for CHD$_2$OH and CH$_3$OD 
but it overestimates the ratio for CH$_2$DOH by one order of magnitude and 
underestimates the ratios for CD$_3$OH and CD$_3$OD by about one order of 
magnitude as well. 

Although it is not straightforward to predict how the best model fit would 
look like for IRAS4A2, adjusting the initial conditions of the model of 
\citet{Faure15} to the D/H ratio that we derived from CH$_2$DOH in this source
(1.2\% instead of 12\%) would probably lead to a reduction of all predicted 
ratios by about one order of magnitude, implying that the model would 
underestimate the abundances of CD$_3$OH and CD$_3$OD by about two orders of 
magnitude.  We conclude from this comparison that either hydrogen isotope 
exchanges between water and methanol during the warm-up phase cannot explain 
the deuteration pattern observed for methanol in IRAS4A2 or that the 
deuterated isotopologs were not statistically distributed at the end of the 
prestellar phase. Measurements in prestellar cores show indeed that the 
distribution of deuterated isotopologs of methanol is not statistical, at 
least in the gas phase after nonthermal desorption \citep[][]{Lin23}.

\begin{figure}[!t]
\centerline{\resizebox{1.0\hsize}{!}{\includegraphics[angle=0]{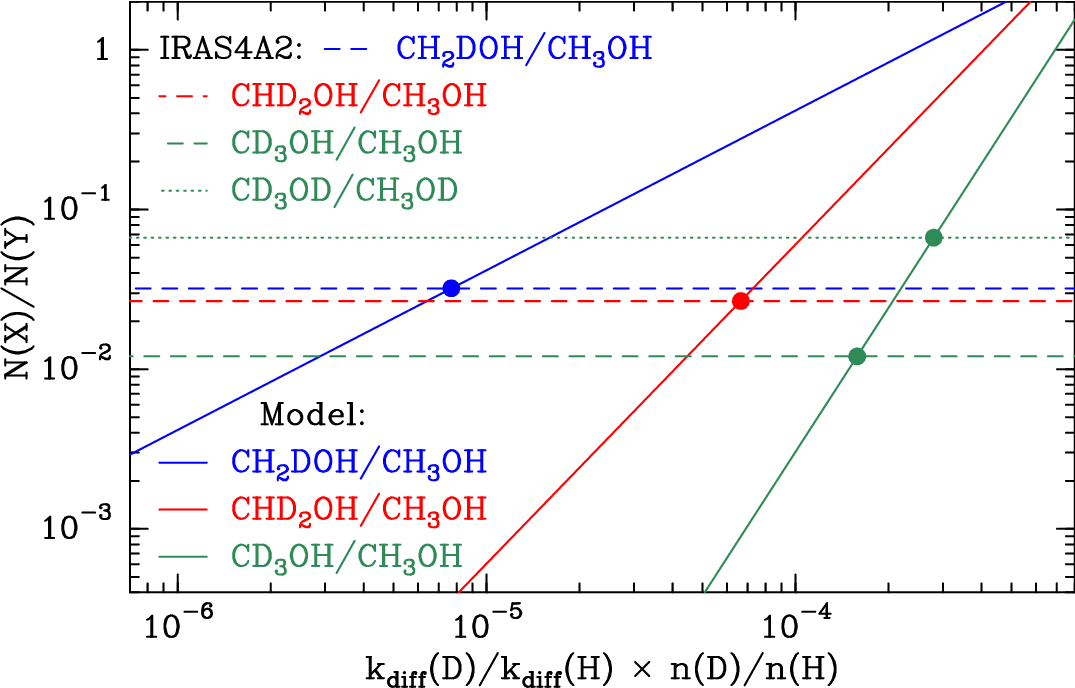}}}
\caption{Comparison of column density ratios measured toward IRAS4A2 (dashed 
lines; using $^{13}$CH$_3$OH as a proxy for methanol) to predictions of the 
simple kinetic model of \citet{Cooper19} (solid 
lines) for CH$_2$DOH (blue), CHD$_2$OH (red), and CD$_3$OH (green) relative to 
CH$_3$OH. The x-axis represents the product of the ratio of the diffusion
constants of atomic D and H with the abundance ratio of D and H on the surface 
of dust grains. The green dotted line represents the column density ratio of 
CD$_3$OD with respect to CH$_3$OD measured toward IRAS4A2. The intersections 
of the observational measurements with the model curves are highlighted by the
thick dots.}
\label{f:modelC19}
\end{figure}

\citet{Cooper19} presented the results of a simple kinetic model of methanol
deuteration in steady-state at 30~K on the basis of new calculations of the 
rate constants of H and D abstraction reactions from the methyl group of 
methanol. The predicted ratios relative to CH$_3$OH are displayed for 
CH$_2$DOH, CHD$_2$OH, and CD$_3$OH in Fig.~\ref{f:modelC19} as a function of 
the product of the ratio of the diffusion constants of atomic D and H (which 
\citeauthor{Cooper19} estimated to be about 0.1) with the ratio of D and H 
atomic abundances on the grains. The horizontal dashed lines in 
Fig.~\ref{f:modelC19} show our results for IRAS4A2. We also added as a dotted 
line the column density ratio of CD$_3$OD relative to CH$_3$OD, assuming that
it would behave in the same way as CD$_3$OH/CH$_3$OH in the model. The 
intersections of the observed and predicted ratios span a range of more
than one order of magnitude along the x axis. This means that abstraction 
reactions from the methyl group of methanol are not sufficient to explain its
deuteration pattern in IRAS4A2, at least within the framework of this simple 
kinetic model.

\begin{figure}[!t]
\centerline{\resizebox{\hsize}{!}{\includegraphics[angle=0]{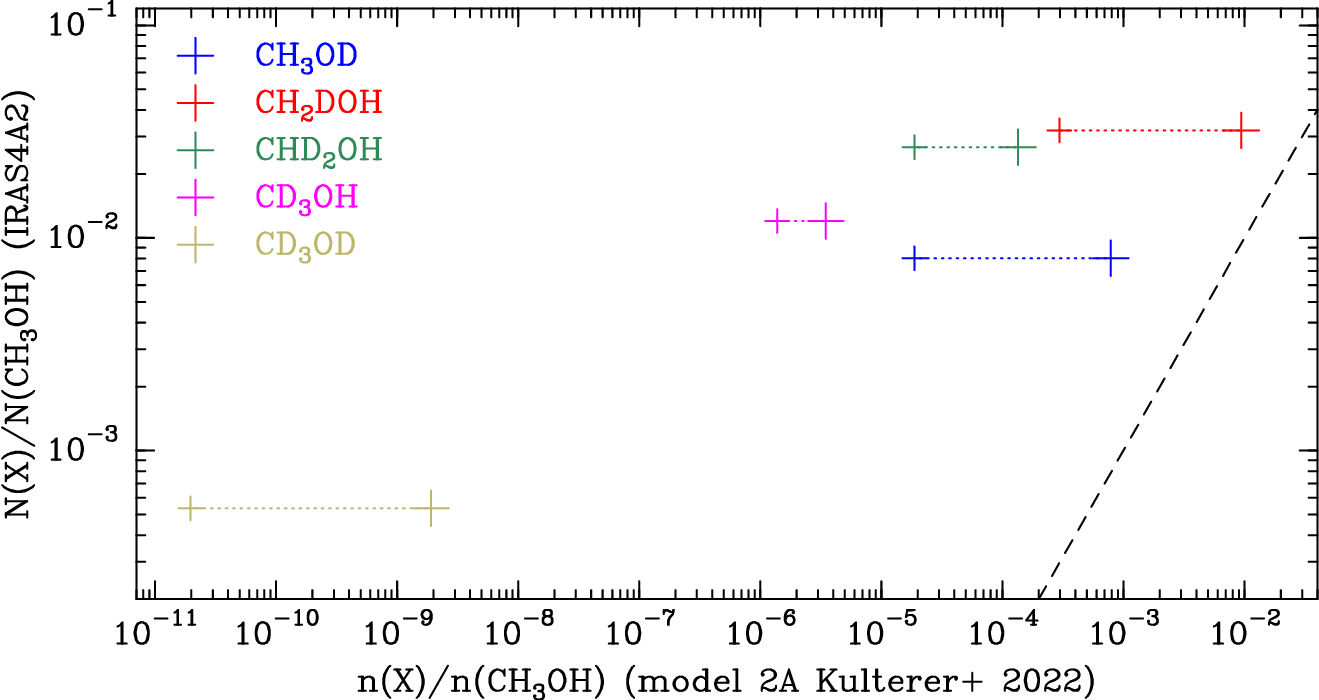}}}
\caption{Column density ratios of deuterated isotopologs relative to
methanol measured toward IRAS4A2 in the gas phase (using $^{13}$CH$_3$OH as a 
proxy for methanol) versus predictions of model 2A of \citet{Kulterer22} on 
the grains in the prestellar phase at an age of $3 \times 10^5$~yr. The small 
and big symbols represent the model results at 10~K and 15~K, respectively. 
The dashed line represents the one-to-one relation.}
\label{f:modelK22}
\end{figure}

A more elaborate gas-grain astrochemical model of a prestellar core at an age
of $3 \times 10^5$~yr was presented by \citet{Kulterer22}. In addition to the 
hydrogenation route of CO leading to methanol, the model includes, like in
\citet{Taquet14}, H and D abstraction and substitution reactions along this 
formation route, informed by the experimental studies mentioned earlier. The 
predicted ratios of grain surface abundances of deuterated isotopologs 
relative to methanol are displayed in Fig.~\ref{f:modelK22} for model 2A of 
\citet{Kulterer22} at two temperatures, 10~K and 15~K, and compared to the 
gas-phase column density ratios that we measured in IRAS4A2. While the model 
at 15~K produces ratios that are only a factor of a few lower than the 
observed ratios for CH$_2$DOH and CH$_3$OD, the abundances of the 
multiply deuterated isotopologs are, like in the layered model of 
\citet{Taquet14}, underestimated by about two orders of  magnitude for
CHD$_2$OH and CD$_3$OH, and even five orders of magnitude for CD$_3$OD, which 
is roughly consistent with the calculations of \citet{Cooper19} (the green 
solid line in Fig.~\ref{f:modelC19} reaches the abscissa of the blue dot at an 
ordinate that lies about four orders of magnitude below the level of the green 
dot). If the ratios measured in IRAS4A2 reflect the deuteration pattern that 
was achieved at the end of the prestellar phase, then a key ingredient leading 
to enhanced abundances of multiply deuterated isotopologs of methanol is 
missing in this prestellar core model. 

\citet{Parise09} presented observational evidence, based on a study of the 
Orion Bar, suggesting that deuterium fractionation of simple molecules 
driven by CH$_2$D$^+$ can occur in the gas phase at warm ($\sim$50~K) 
temperatures. Gas-phase chemistry models of \citet{Millar89} also predicted
that significant deuterium fractionation of simple molecules driven by 
CH$_2$D$^+$, C$_2$HD$^+$, and associated species can occur even at 70~K. 
However, the chemical timescales of this high-temperature deuteration route 
appear to be too long \citep[$>$10$^5$ yr,][]{Rodgers96} to modify the 
deuteration pattern of methanol in the gas phase of hot corinos after thermal 
desorption. 

In summary, all the models mentioned in this section so far end up
underestimating (by several orders of magnitude) the production of multiply
deuterated isotopologs of methanol during low-mass star formation. This is 
further exacerbated by our detection of CD$_3$OD in IRAS4A2.

A more recent gas-grain chemical model of methanol deuteration during the 
prestellar phase was presented by \citet{Riedel23,Riedel25}. Like 
\citet{Taquet14} and \citet{Kulterer22}, the model includes H and D abstraction
and substitution reactions along the CO hydrogrenation route. The model also
accounts for reactive desorption and nondiffusion processes. The model was 
tuned to the physical properties of the prestellar core L1544, in particular 
with a dust temperature varying from $\sim$15~K in the outskirts to $\sim$6~K in
the inner region. \citet{Riedel23,Riedel25} only reported predictions for the 
CH$_2$DOH/CH$_3$OH column density ratio. The same model was used in 
\citet{Spezzano26}, where predictions for both CH$_2$DOH and CHD$_2$OH were 
reported. The model version that includes abstraction reactions (model D5) 
produces gas-phase column density ratios of $\sim$5--7\% for 
CH$_2$DOH/CH$_3$OH, slightly higher but comparable to what we measured in 
IRAS4A2, and $\sim$10\% for CHD$_2$OH/CH$_2$DOH, which is nearly one order of 
magnitude lower than what we measured in IRAS4A2. If the distribution of
deuterated isotopologs in the model is similar for both the gas phase and the 
solid phase then, like for the other models discussed in this section, we 
conclude that this model does not reproduce the high abundance of 
doubly deuterated methanol relative to the singly deuterated isotopolog in 
IRAS4A2. It would be interesting to know what the model predicts for the ice
mantles at the end of the prestellar phase, in particular for the more 
deuterated isotopologs CD$_3$OH and CD$_3$OD.

Given that formaldehyde is an intermediate species in the hydrogenation
scheme that leads to methanol, we briefly discuss the 
deuteration pattern of H$_2$CO. In their study of L1544, 
\citet{Spezzano22} reported that, for both H$_2$CO and H$_2$CS, their chemical
model fails to reproduce the column density ratio of the doubly to 
singly deuterated isotopologs while it accounts well for the ratio of the
singly deuterated species to the main isotopolog. This led them to conclude 
that the reaction network for the formation of D$_2$CO and D$_2$CS is not 
complete yet. In IRAS16293B, \citet{Persson18} reported HDCO/H$_2$CO and 
D$_2$CO/HDCO column density ratios of 6.5\% and 13\%, respectively. The former 
ratio is close to the prediction of \citet{Taquet14}'s model at the beginning 
of the Class 0 phase (4.1\%), but the latter ratio is a factor of $\sim$3 
higher than predicted by the model (4.4\%). \citet{Chahine24} reported maps of 
HDCO and D$_2$CO toward IRAS4A. While they focused their analysis on the 
outflow cavity walls on larger scales, their maps indicate a D$_2$CO/HDCO 
intensity ratio of $\sim$0.6 toward IRAS4A2, which translates into a 
D$_2$CO/HDCO column density ratio of $\sim$70\% if we assume optically thin 
emission and a temperature of 150--200~K.  This value is much higher than the
prediction of \citet{Taquet14}'s model.
Formaldehyde and its isotopologs are readily detected in the COMPASS data of 
IRAS4A2. However, the derivation of its deuteration levels for comparison with 
the methanol D/H ratios presented here is beyond the scope of this article. 
Formaldehyde and its isotopologs display extended and compact emission as 
demonstrated already by the maps published in \citet{Chahine24}. Our 
preliminary analysis of the compact emission suggests that the D/H ratio in 
singly and doubly deuterated formaldehyde is on par with the trends seen here
in methanol for IRAS4A2 and with the formaldehyde results obtained for 
IRAS16293B \citep[][]{Persson18}. The evidence available thus far suggests 
that, akin to the case of methanol, current chemical models fail to reproduce 
the high abundance of doubly deuterated formaldehyde.

\section{Conclusions}
\label{s:conclusions}

We used the spectral line survey performed toward the low-mass protostar 
IRAS4A2 within the framework of the ALMA large program COMPASS to study the 
deuteration pattern of methanol in the gas phase of the hot corino. All 
deuterated isotopologs of methanol for which spectroscopic predictions are 
available in public databases were detected. In particular, we report the 
first robust interstellar detection of the fully deuterated isotopolog of 
methanol, CD$_3$OD. This represents the first identification of a 
quadruply deuterated molecule in the interstellar medium.
In light of this detection, it is highly likely that the other 
deuterated isotopologs of methanol (CH$_2$DOD and CHD$_2$OD) will be easily 
detected in IRAS4A2 as soon as spectroscopic predictions become available for 
these species.

Our LTE modeling of the IRAS4A2 spectra yields unrealistically low column 
density ratios of methanol to its $^{13}$C and $^{18}$O isotopologs, while both 
the ratio of the $^{13}$C and $^{18}$O isotopologs and the ratio of the $^{18}$O 
and $^{17}$O isotopologs are consistent with the isotopic ratios expected for 
the local ISM. This suggests that the column density that we derived for 
methanol is underestimated. The methanol emission is barely resolved in our 
data set. While we managed to constrain its size by 2D Gaussian 
fits to integrated intensity maps, one possible explanation for our 
underestimate of the column density of methanol is that its emission on scales 
smaller than the beam is clumpy rather than monolithic.

The deuteration pattern of methanol in IRAS4A2 is similar to the pattern that
was measured in several other low-mass protostars, with an increase of the
apparent D/H ratio, computed assuming a statistical distribution of D and H, 
from the singly to multiply deuterated isotopologs. The comparison of this
observed pattern to the predictions of astrochemical models of prestellar and
protostellar cores focused on deuteration published in the past
decade indicates that none of the models quantitatively reproduces the high 
deuteration level of the multiply deuterated isotopologs, which is further
exacerbated by the high abundance of CD$_3$OD relative to
methanol ($\sim$4$\times 10^{-4}$). A key ingredient promoting the formation of 
multiply deuterated isotopologs of methanol seems to be missing in these models.

\section*{Data availability}
Appendix B is available on Zenodo via https://doi.org/10.5281/zenodo.19254994.

\begin{acknowledgements}
This paper makes use of the following ALMA data: ADS/JAO.ALMA\#2022.1.00316.L. 
ALMA is a partnership of ESO (representing its member states), NSF (USA) and 
NINS (Japan), together with NRC (Canada), NSTC and ASIAA (Taiwan), and KASI 
(Republic of Korea), in cooperation with the Republic of Chile. The Joint ALMA 
Observatory is operated by ESO, AUI/NRAO and NAOJ.
The National Radio Astronomy Observatory is a facility of the National Science
Foundation operated under cooperative agreement by Associated Universities,
Inc.
The project that led to this publication has received support from ORP that is 
funded by the European Union's Horizon 2020 research and innovation programme 
under grant agreement No 101004719 [ORP]. 
We acknowledge financial support from the following: the European Research 
Council (ERC) under the European Union's Horizon 2020 research and innovation 
programme (ERC Starting Grant ``Chemtrip'', grant agreement No 949278; A.C.,
F.C.-S.M.);
the National Research Foundation of Korea (NRF), funded by the Ministry of 
Science and ICT (MSIT) under grant numbers 2021R1A2C1011718 and 
RS-2024-00416859 (J.-E.L., J.-H.J., H.-S.Y.),
and by the Ministry of Education through the Basic Science Research Program 
under grant number RS-2025-25402622 (J.-H.J.); 
the Thematic Action ``Physique et Chimie du Milieu Interstellaire'' (PCMI) of 
INSU Programme National ``Astro'', with contributions from CNRS Physique \& 
CNRS Chimie, CEA, and CNES (S.M.); 
the Ministry of Education of Taiwan (a grant from Center for Informatics and 
Computation in Astronomy, CICA, and grant number 110J0353I9) and the National 
Science and Technology Council, Taiwan (Grant NSTC111-2112-M-007-014-MY3, 
NSTC113-2639-M-A49-002-ASP, and NSTC113-2112-M-007-027; D.H.); 
the Max Planck Society (S.S.); 
the ESO and IAU Gruber Foundation Fellowship programs (P.N.); 
the Ministry of Education, Culture, Sports, Science, and Technology of Japan 
(``Grant-in-Aid''; 20H05845, 20H05844) and RIKEN (pioneering project 
``Evolution of Matter in the Universe''; (Y.-L.Y.); 
RIKEN through the Special Postdoctoral Researcher Program (Fellowships; S.Z.).

\end{acknowledgements}

\bibliographystyle{aa}
\bibliography{aa59642-26}

\begin{appendix}
\label{appendix}

\clearpage
\section{Population diagrams, sizes, lines}
\label{a:popdiag_sizes}

\begin{figure}[!h]
\centerline{\resizebox{\hsize}{!}{\includegraphics[angle=0]{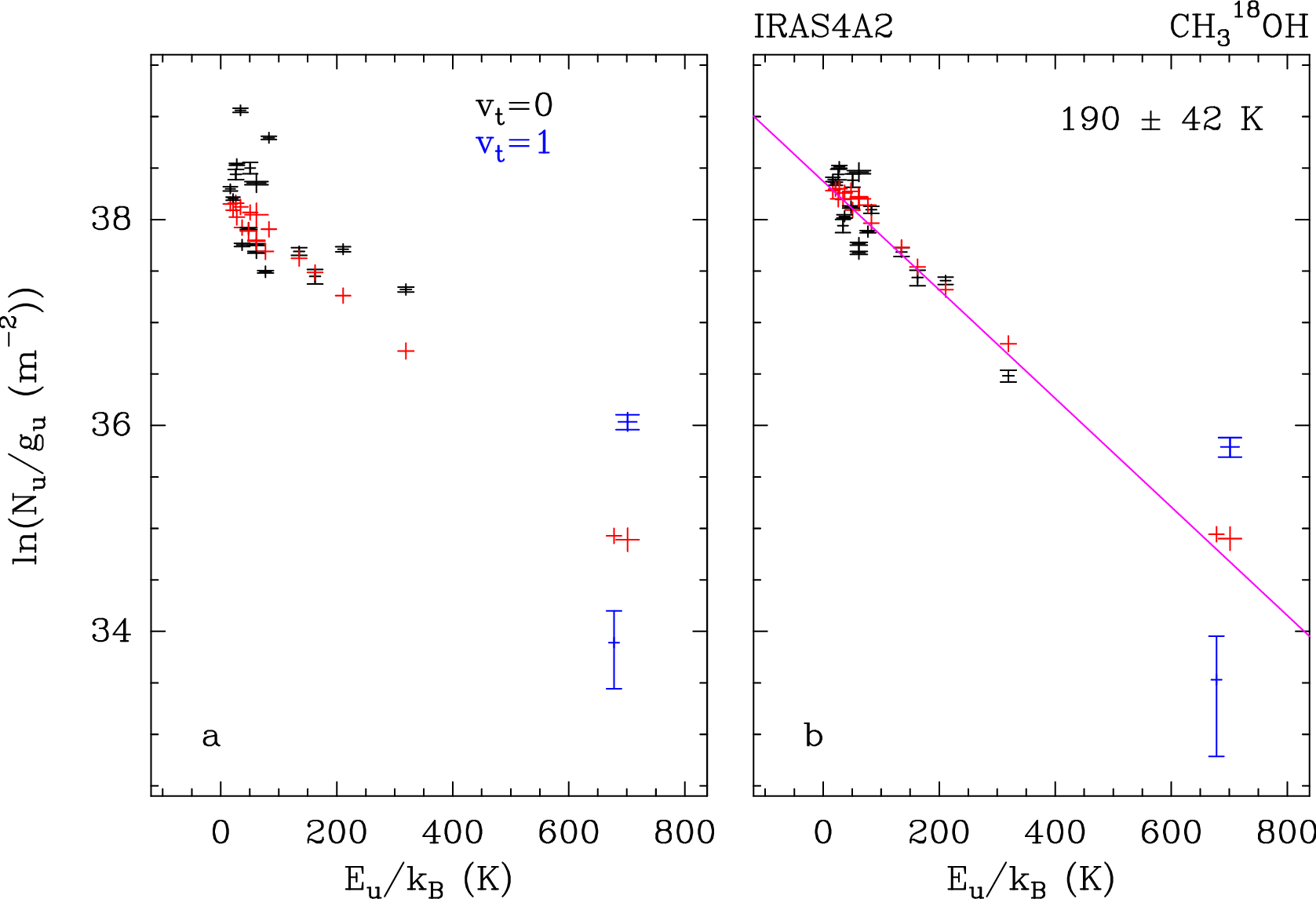}}}
\vspace*{-1ex}
\caption{Population diagram of CH$_3$$^{18}$OH toward IRAS4A2. See caption of
Fig.~\ref{f:popdiags} for details.}
\label{f:popdiag_ch3oh_18o}
\end{figure}

\begin{figure}[!h]
\centerline{\resizebox{\hsize}{!}{\includegraphics[angle=0]{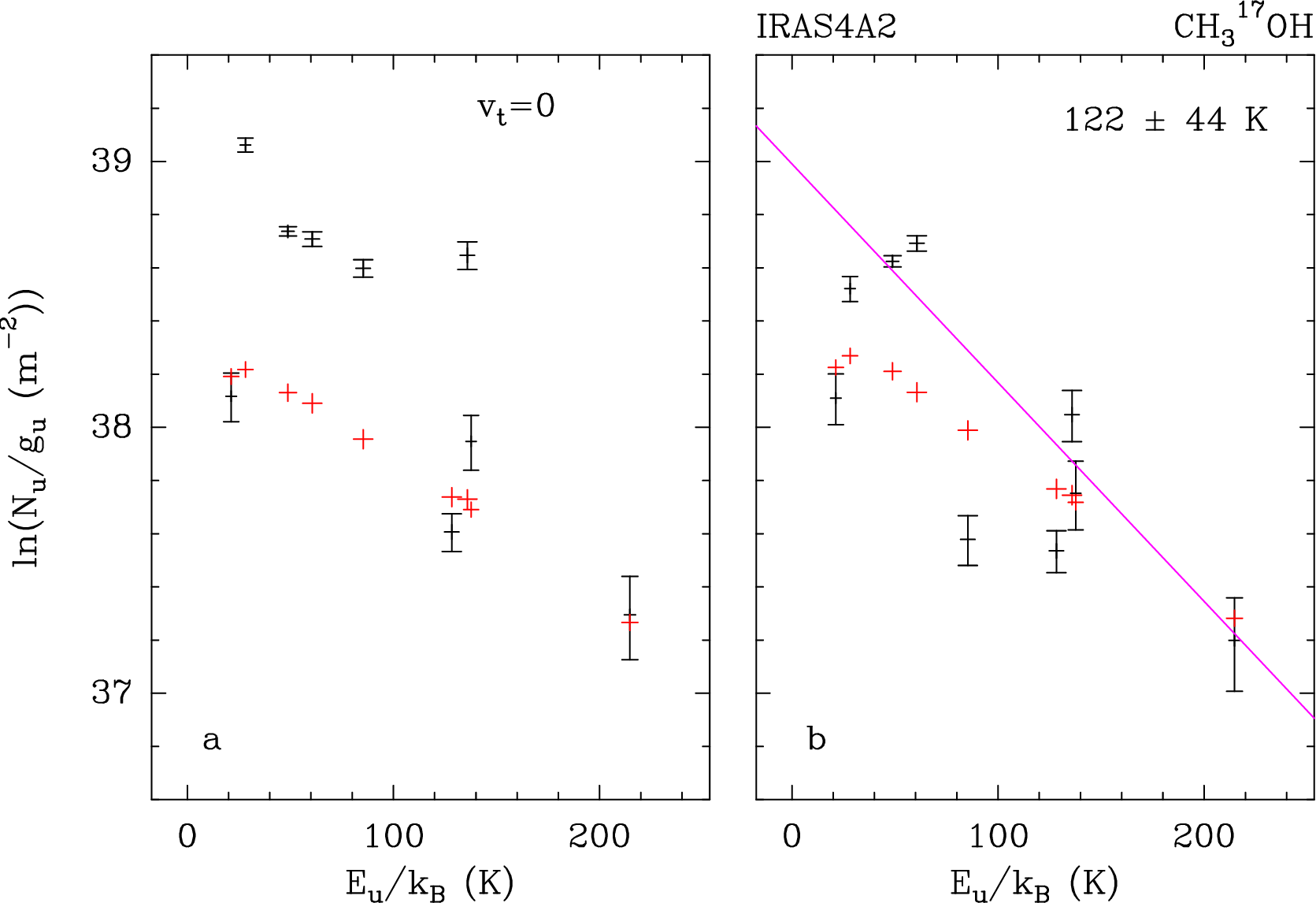}}}
\vspace*{-1ex}
\caption{Same as Fig.~\ref{f:popdiag_ch3oh_18o}, but for CH$_3$$^{17}$OH.}
\label{f:popdiag_ch3oh_17o}
\end{figure}

\begin{figure}[!h]
\centerline{\resizebox{\hsize}{!}{\includegraphics[angle=0]{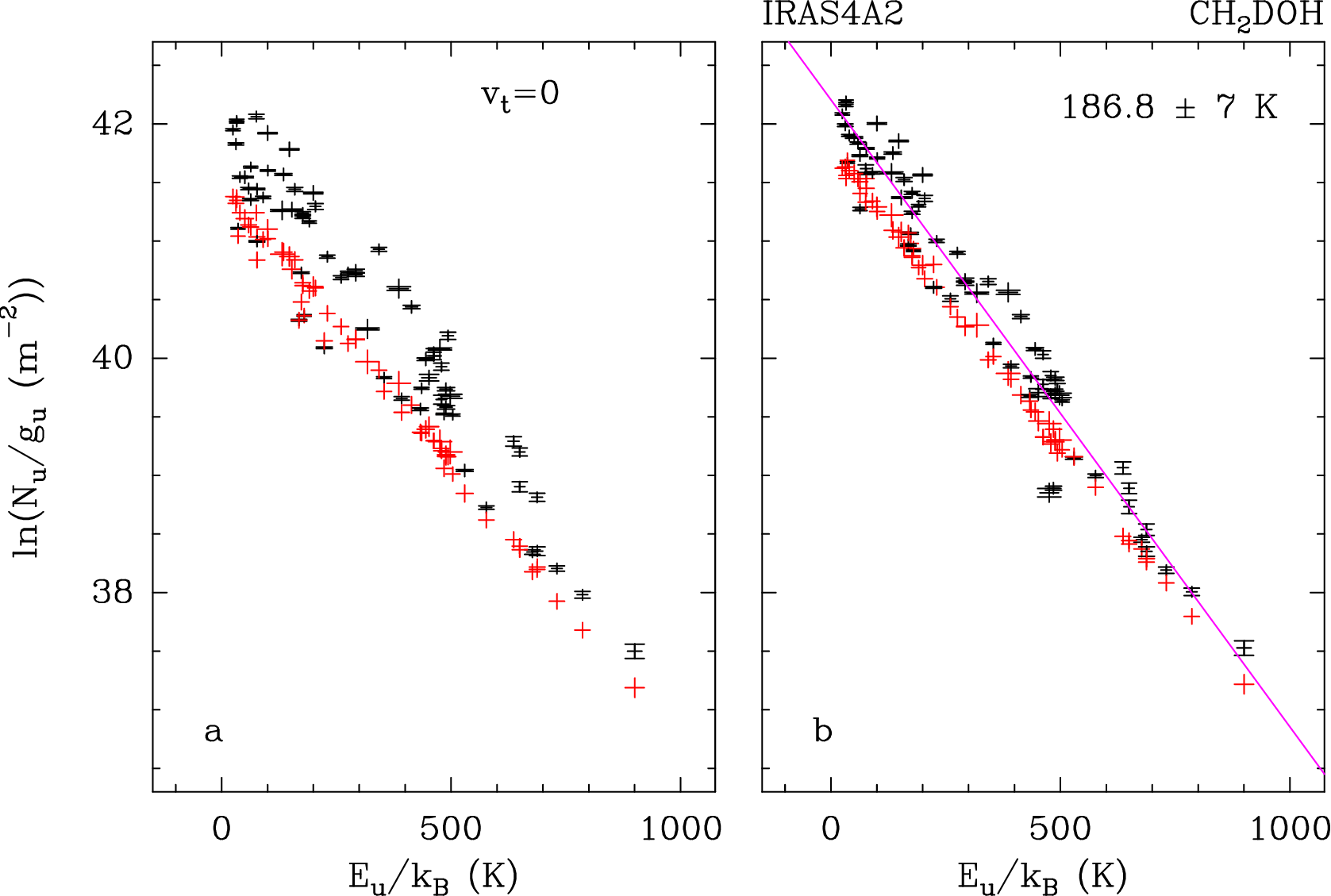}}}
\vspace*{-1ex}
\caption{Same as Fig.~\ref{f:popdiag_ch3oh_18o}, but for CH$_2$DOH.}
\label{f:popdiag_ch2doh}
\end{figure}

\begin{figure}[!h]
\centerline{\resizebox{\hsize}{!}{\includegraphics[angle=0]{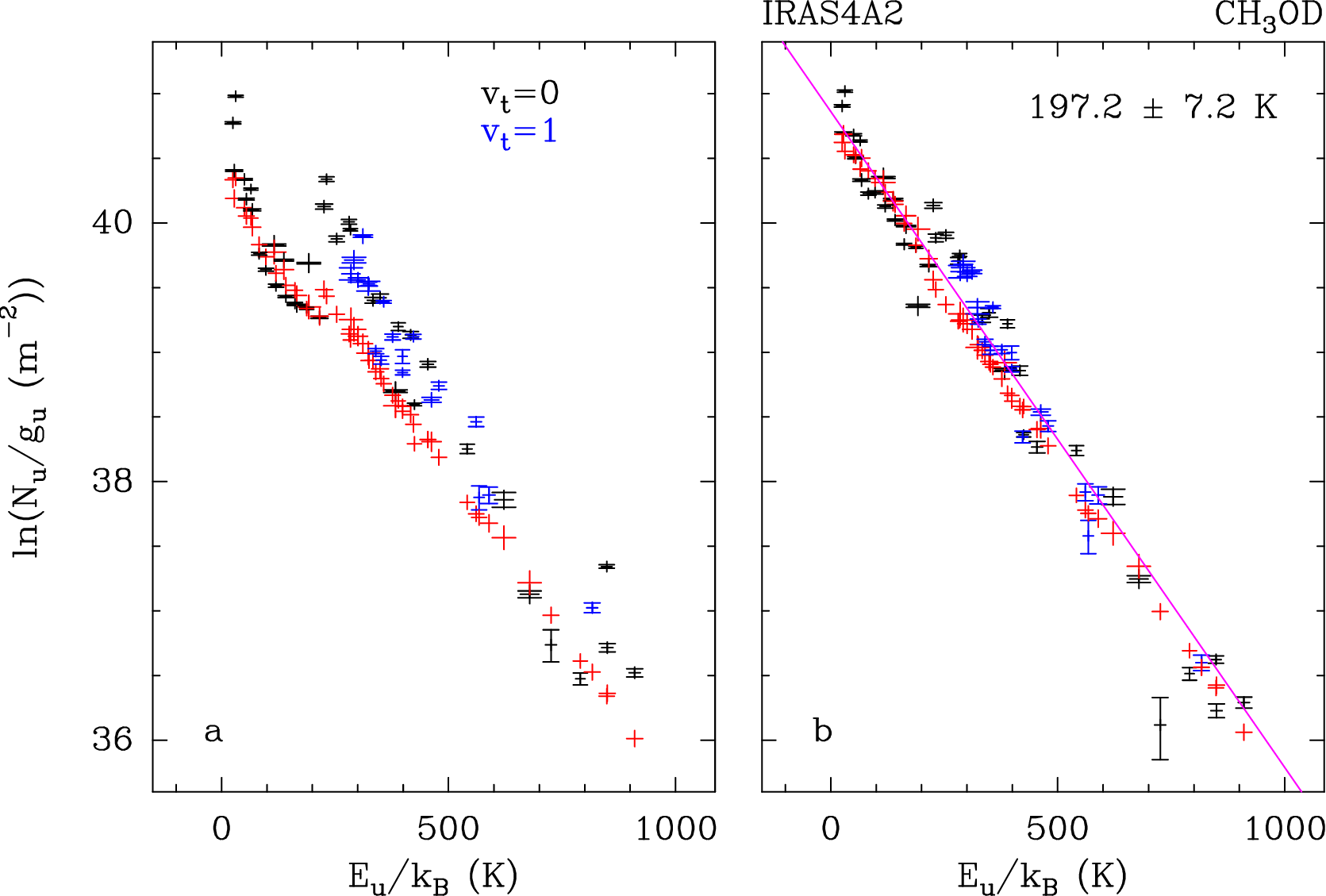}}}
\vspace*{-1ex}
\caption{Same as Fig.~\ref{f:popdiag_ch3oh_18o}, but for CH$_3$OD.}
\label{f:popdiag_ch3od}
\end{figure}

\begin{figure}[!h]
\centerline{\resizebox{\hsize}{!}{\includegraphics[angle=0]{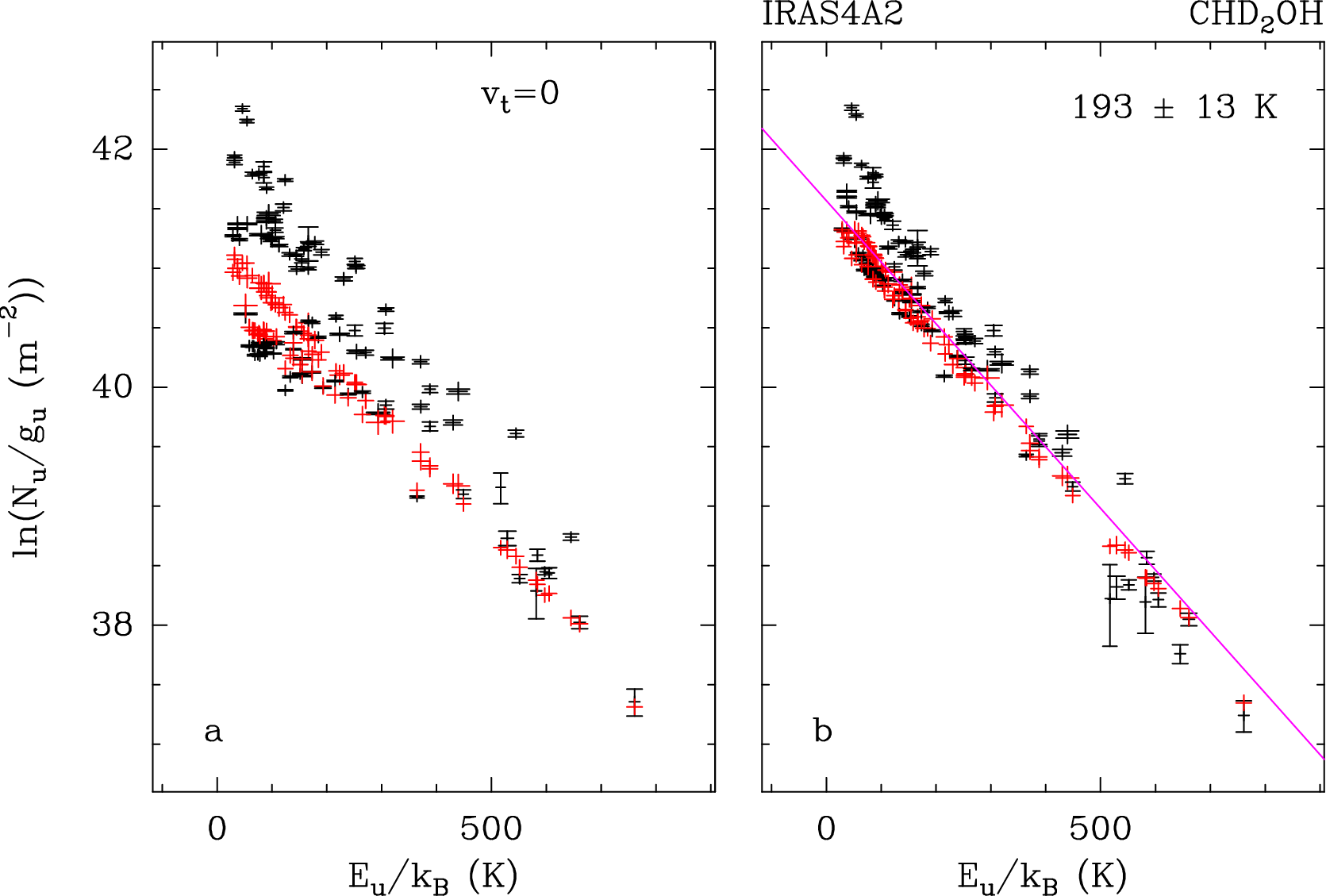}}}
\vspace*{-1ex}
\caption{Same as Fig.~\ref{f:popdiag_ch3oh_18o}, but for CHD$_2$OH.}
\label{f:popdiag_chd2oh}
\end{figure}

\begin{figure}[!h]
\centerline{\resizebox{\hsize}{!}{\includegraphics[angle=0]{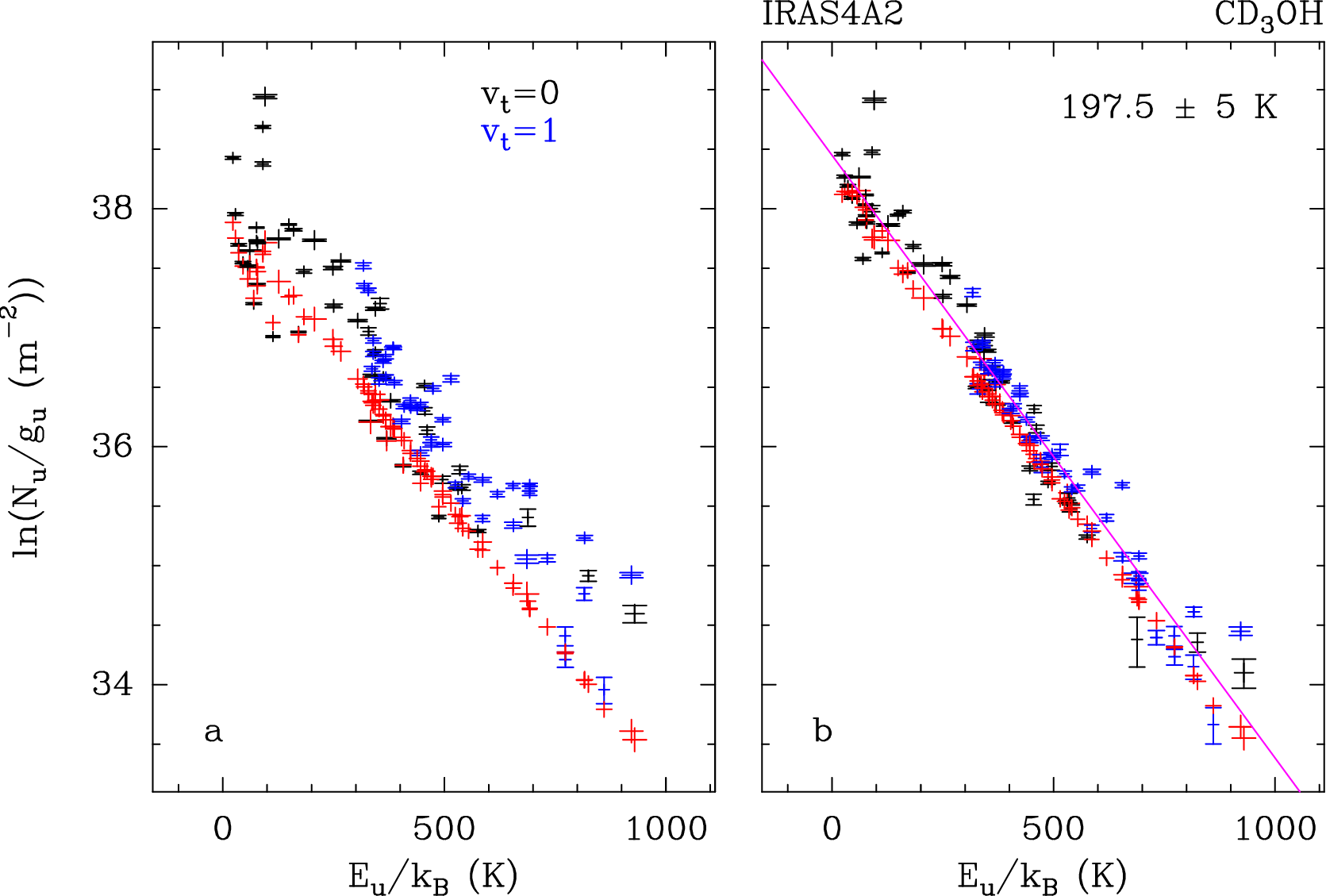}}}
\vspace*{-1ex}
\caption{Same as Fig.~\ref{f:popdiag_ch3oh_18o}, but for CD$_3$OH.}
\label{f:popdiag_cd3oh}
\end{figure}

\begin{figure*}
\begin{center}
\resizebox{0.76\hsize}{!}{\includegraphics[angle=0]{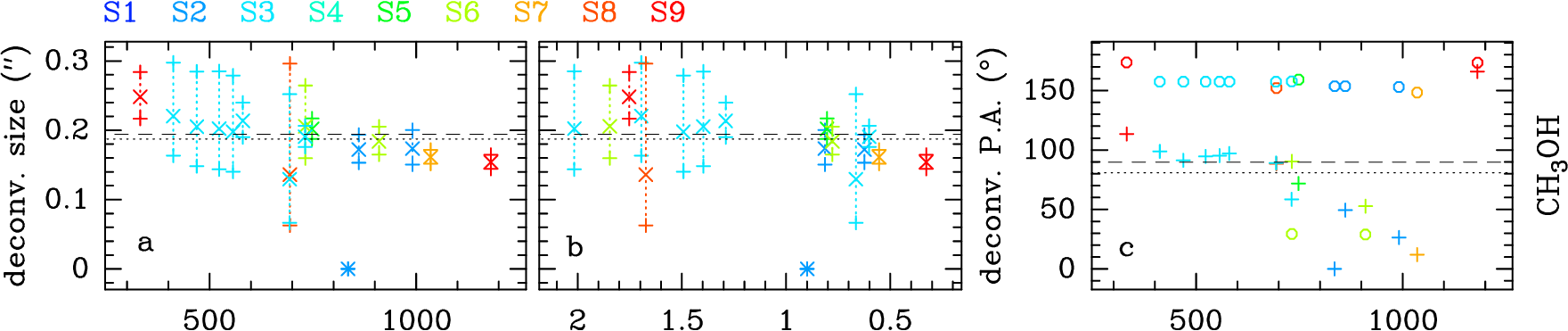}}\\[0.7ex]
\resizebox{0.76\hsize}{!}{\includegraphics[angle=0]{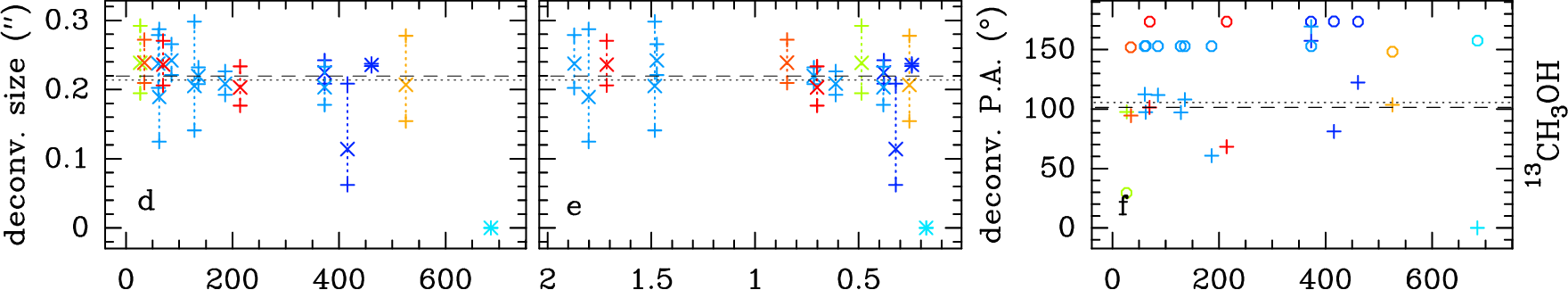}}\\[0.7ex]
\resizebox{0.76\hsize}{!}{\includegraphics[angle=0]{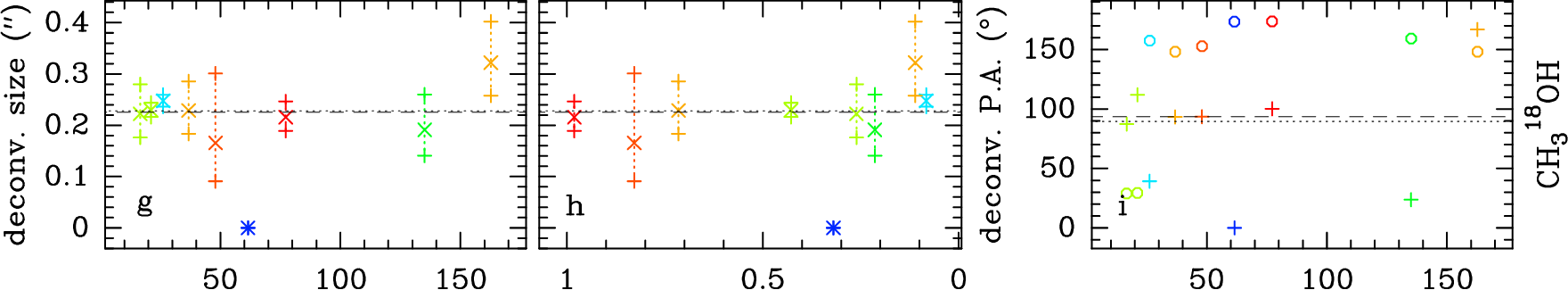}}\\[0.7ex]
\resizebox{0.76\hsize}{!}{\includegraphics[angle=0]{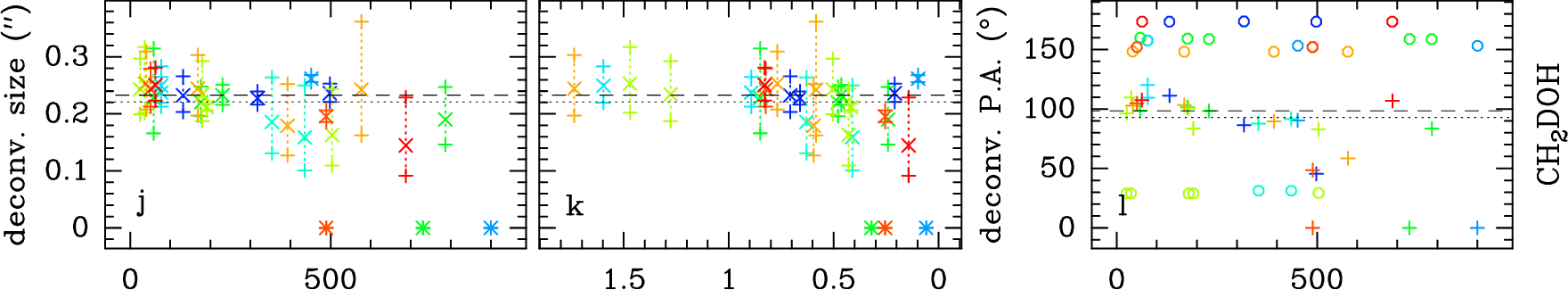}}\\[0.7ex]
\resizebox{0.76\hsize}{!}{\includegraphics[angle=0]{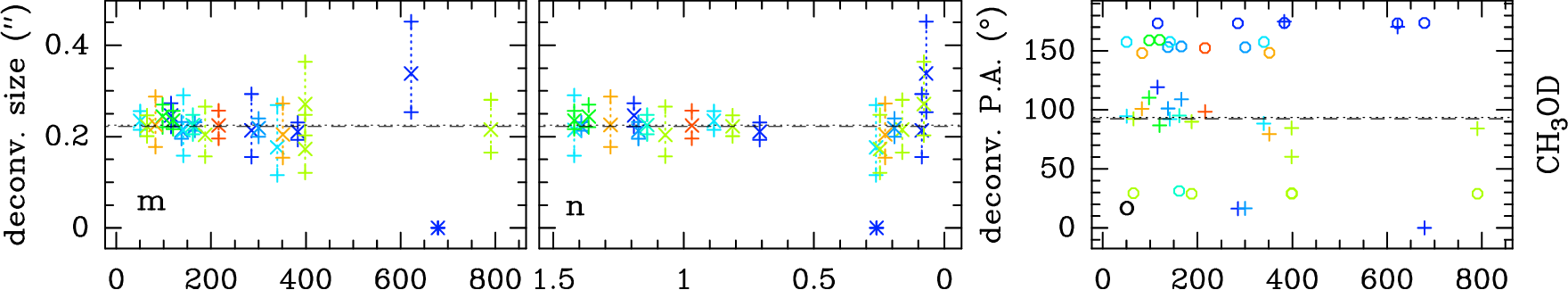}}\\[0.7ex]
\resizebox{0.76\hsize}{!}{\includegraphics[angle=0]{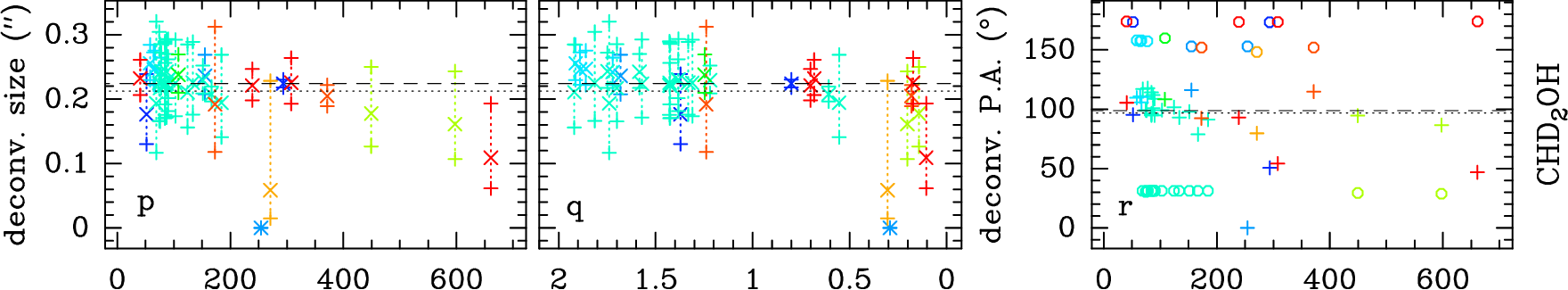}}\\[0.7ex]
\resizebox{0.76\hsize}{!}{\includegraphics[angle=0]{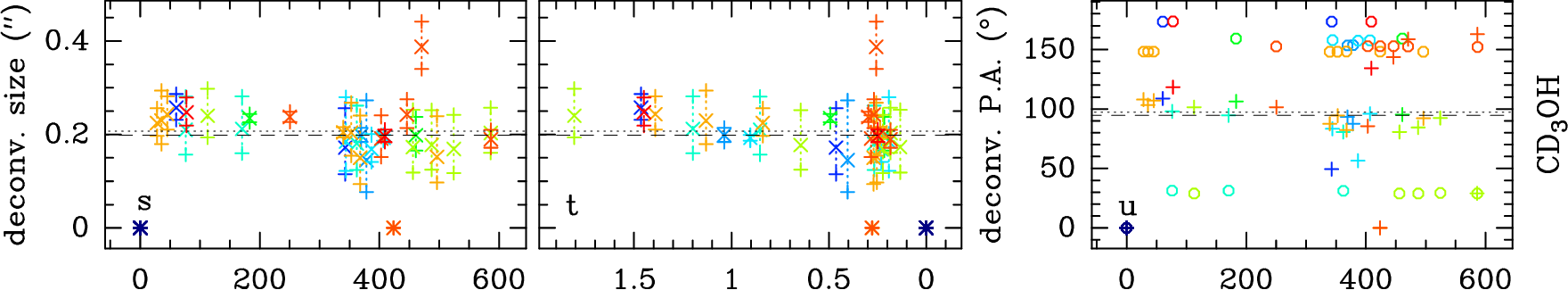}}\\[0.7ex]
\resizebox{0.76\hsize}{!}{\includegraphics[angle=0]{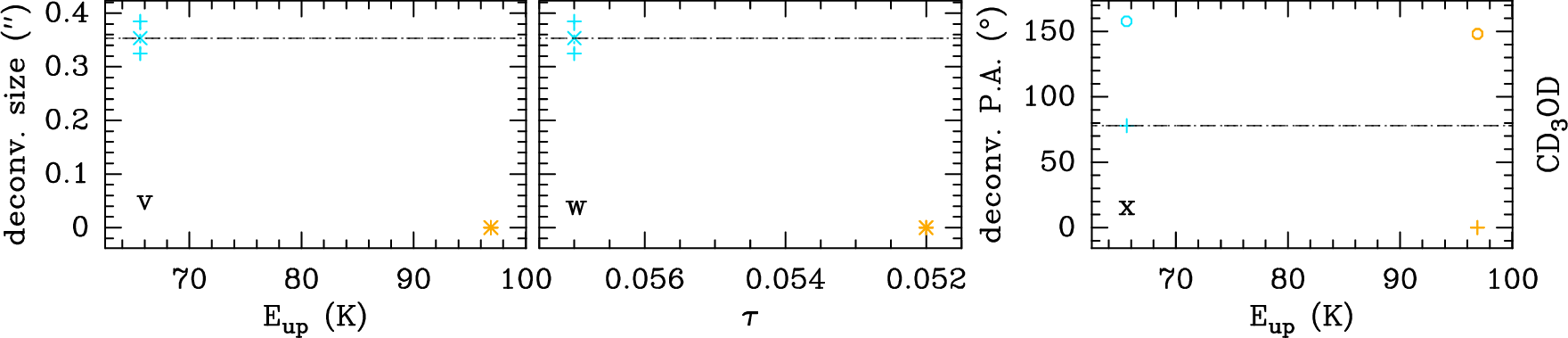}}
\end{center}
\vspace*{-3.5ex}
\caption{Emission sizes of methanol and seven of its isotopologs (one per row)
as a function of upper-level energy (left column) and line opacity (middle 
column), and position angle as a function of upper-level energy (right column). 
The colors indicate the spectral setup (S1--S9) in which each transition was 
observed. In the left and middle columns, the plus symbols of each transition
represent the deconvolved major and minor axes of its emission and the cross 
in-between is their geometrical average. The dashed and dotted lines show the 
median and mean of the geometrical averages, respectively. In the right 
column, each plus symbol represents the position angle (east from the north) 
of the emission and the circle at the same abscissa value indicates the 
position angle of the beam. The dashed and dotted lines represent the median 
and mean of the position angles of the emission, respectively. Cases for which 
the deconvolution failed are set to 0 and were ignored for the calculation of 
the median and mean.}
\label{f:sizes}
\end{figure*}

Figures~\ref{f:popdiag_ch3oh_18o}--\ref{f:popdiag_cd3oh} show the population
diagrams of CH$_3$$^{18}$OH,  CH$_3$$^{17}$OH, CH$_2$DOH, CH$_3$OD, CHD$_2$OH,
and CD$_3$OH, respectively, toward IRAS4A2. Figure~\ref{f:sizes} displays
the deconvolved sizes and position angles of the emission of methanol and seven
of its isotopologs. Table~\ref{t:cd3od_det_int} lists the spectroscopic
parameters and intensities of the detected lines of CD$_3$OD.

\vspace*{-1ex}

\vspace*{-1ex}

\vspace*{-1ex}

\clearpage

\clearpage
\begin{table*}
 \begin{center}
 \caption{
 Spectroscopic parameters and integrated intensities of transitions of CD$_3$OD detected toward IRAS4A2 in the COMPASS survey.
}
 \label{t:cd3od_det_int}
 \vspace*{0.0ex}
 \begin{tabular}{lrcrcccrccc}
 \hline\hline
 \multicolumn{1}{c}{Transition\tablefootmark{(a)}} & \multicolumn{1}{c}{Frequency\tablefootmark{(b)}}  & \multicolumn{1}{c}{$A_{\rm ul}$\tablefootmark{(c)}} & \multicolumn{1}{c}{$E_{\rm u}$\tablefootmark{(d)}}  & \multicolumn{1}{c}{$g_{\rm u}$\tablefootmark{(e)}} & \multicolumn{1}{c}{$I_{\rm obs}$\tablefootmark{(f)}} & \multicolumn{1}{c}{$I_{\rm mod}$\tablefootmark{(g)}} & \multicolumn{1}{c}{$I_{\rm all}$\tablefootmark{(h)}} & \multicolumn{1}{c}{$\underline{S_{\rm obs}}$\tablefootmark{(i)}} & \multicolumn{1}{c}{$\underline{S_{\rm dec}}$\tablefootmark{(j)}} & \multicolumn{1}{c}{$\underline{S_{\rm mod}}$\tablefootmark{(k)}} \\ 
 \multicolumn{1}{c}{$S J_{K_a,K_c}$} & \multicolumn{1}{c}{\small (MHz)} & \multicolumn{1}{c}{\small (10$^{-5}$ s$^{-1}$)} & \multicolumn{1}{c}{\small (K)} & & \multicolumn{1}{c}{\small (K km s$^{-1}$)} & \multicolumn{2}{c}{\small (K km s$^{-1}$)} & \multicolumn{1}{c}{\hspace*{-2ex}$N$} & \multicolumn{1}{c}{\hspace*{-2ex}$N$} & \multicolumn{1}{c}{\hspace*{-2ex}$N$} \\ 
 \hline \\[-1.8ex]
E 8$_{0,8}$ -- 7$_{-1,7}$ & 288875.604(2) &  12.1 &    65.6 & 272 &   5.5(5) &   4.8 &   5.1 &    11 &    11 &   9.8 \\ 
E 19$_{-4,16}$ -- 19$_{-3,17}$ & 289772.199(1) &  13.9 &   372.7 & 624 &   4.9(4) &   2.6 &   3.5 &    12 &   9.7 &   6.2 \\ 
E 13$_{-4,10}$ -- 13$_{-3,11}$ & 291656.386(1) &  13.5 &   197.9 & 432 &   6.8(4) &   4.7 &   4.8 &    18 &    17 &    12 \\ 
E 11$_{-4,8}$ -- 11$_{-3,9}$ & 291867.441(1) &  13.1 &   153.7 & 368 &   9.6(4) &   4.8 &  10.7 &    25 &   9.6 &    12 \\ 
E 9$_{-4,6}$ -- 9$_{-3,7}$ & 291982.995(1) &  12.5 &   116.6 & 304 &   8.3(4) &   4.5 &  10.0 &    22 &   7.4 &    12 \\ 
E 8$_{-4,5}$ -- 8$_{-3,6}$ & 292015.994(1) &  11.9 &   100.7 & 272 &   5.8(4) &   4.2 &   4.4 &    15 &    15 &    11 \\ 
E 8$_{-1,8}$ -- 7$_{-1,7}$ & 293037.062(1) &  9.79 &    65.8 & 272 &   6.3(5) &   4.3 &   5.0 &    13 &    12 &   9.0 \\ 
E 8$_{-5,4}$ -- 7$_{-5,3}$ & 294365.802(1) &  6.52 &   119.6 & 272 &   6.8(7) &   4.6 &   4.8 &    10 &    10 &   7.0 \\ 
E 8$_{-3,6}$ -- 7$_{-3,5}$ & 294366.149(1) &  8.95 &    86.7 & 272 & -- & -- & -- & -- & -- & -- \\ 
E 8$_{1,7}$ -- 7$_{1,6}$ & 295497.606(1) &  10.4 &    68.6 & 272 &   4.4(7) &   3.9 &   4.6 &   6.6 &   5.5 &   5.9 \\ 
A 10$_{0,10}$ -- 9$_{1,8}$ & 301754.918(1) &  16.5 &    96.9 & 231 &   5.8(7) &   4.6 &   4.9 &   8.7 &   8.3 &   6.9 \\ 
 \hline
 \end{tabular}
 \end{center}
 \vspace*{-2.5ex}
 \tablefoot{
 \tablefoottext{a}{$S$ corresponds to the symmetry (A or E). See \citet{Ilyushin23} for more details about the quantum numbers.}
 \tablefoottext{b}{The value in parentheses represents the frequency uncertainty in unit of the last digit.}
 \tablefoottext{c}{Einstein coefficient for spontaneous emission.}
 \tablefoottext{d}{Upper-level energy.}
 \tablefoottext{e}{Upper-level degeneracy.}
 \tablefoottext{f}{Integrated intensity of the observed spectrum in brightness temperature scale. The statistical standard deviation is given in parentheses in unit of the last digit.}
 \tablefoottext{g}{Integrated intensity of the best-fit synthetic spectrum of CD$_3$OD.}
 \tablefoottext{h}{Integrated intensity of the model that contains the contribution of all identified molecules, including CD$_3$OD.}
 \tablefoottext{i}{Signal-to-noise ratio of $I_{\rm obs}$.}
 \tablefoottext{j}{Signal-to-noise ratio of the integrated intensity, $I_{\rm dec}$, decontaminated from the contribution of molecules other than CD$_3$OD, that is, $I_{\rm dec} = I_{\rm obs}-(I_{\rm all}-I_{\rm mod})$, computed with the uncertainty of $I_{\rm obs}$.}
 \tablefoottext{k}{Signal-to-noise ratio of $I_{\rm mod}$ computed with the uncertainty of $I_{\rm obs}$. In the last six columns, a value followed by a dash in the next row represents the value obtained for a group of transitions that are not resolved in the astronomical spectrum.}
 }
 \end{table*}

\end{appendix}

\end{document}